\documentclass{llncs}
\usepackage{times}
\usepackage{bbm}
\usepackage{latexsym}
\usepackage{amssymb}
\usepackage{amsmath}
\usepackage{amstext}
\usepackage{theorem}
  \theorembodyfont{\upshape}
\usepackage{styles/proof}
\usepackage[dvips]{graphicx}
\usepackage{pst-node}
\usepackage{psfrag}
\usepackage{xspace}
\newcommand{\andl}{\wedge}
\newcommand{\notl}{\neg}

\newcommand{\true}{\textsc{true}}
\newcommand{\false}{\textsc{false}}

\newcommand{\uP}{\ensuremath{\mathcal{U}}}

\newcommand{\eP}{\ensuremath{\mathcal{E}}}
\newcommand{\sP}{\ensuremath{\mathcal{S}}}
\newcommand{\iP}{\ensuremath{\mathcal{I}}}
\newcommand{\xP}{\ensuremath{\mathcal{X}}}

\newcommand{\mU}{\ensuremath{\mathcal{U}}}
\newcommand{\mT}{\ensuremath{\mathcal{T}}}

\newcommand{\mL}{\ensuremath{\mathcal{L}}}
\newcommand{\mA}{\ensuremath{\mathcal{A}}}
\newcommand{\mB}{\ensuremath{\mathcal{B}}}
\newcommand{\mC}{\ensuremath{\mathcal{C}}}
\newcommand{\mF}{\ensuremath{\mathcal{F}}}
\newcommand{\mG}{\ensuremath{\mathcal{G}}}

\newcommand{\mR}{\ensuremath{\mathcal{R}}}

\newcommand{\gI}{\mathfrak{I}}

\newcommand{\gM}{\mathfrak{M}}
\newcommand{\ga}{\mathfrak{a}}

\newcommand{\gA}{\mathfrak{V}}

\newcommand{\gC}{\mathfrak{C}}

\newcommand{\gJ}{\mathfrak{J}}

\newcommand{\auf}{\left\langle}
\newcommand{\zu}{\right\rangle}
\newcommand{\Nat}{\mathbbm{N}}

\newcommand{\sometime}{\lozenge}
\newcommand{\sometimes}{\lozenge}

\newcommand{\until}{\mathrm{U}}
\newcommand{\unless}{\mathrm{W}}

\newcommand{\union}{\cup}

\def\true{\hbox{\bf true}}
\def\false{\hbox{\bf false}}

\newcommand{\wfL}{\ensuremath{\mathit{waitforL}}}
\newcommand{\TProb}{\textsf{P}}
\newcommand{\XTProb}{\textsf{XP}}

\renewcommand{\implies}{\Rightarrow}
\newcommand{\EAAE}{\ensuremath{\exists^*\forall^2\exists^*}\xspace}

\newcommand*{\fotl}{\ensuremath{\mathsf{FOTL}}\xspace}
\newcommand{\Grounded}{Derived\xspace}
\newcommand{\grounded}{derived\xspace}
\newcommand{\moreboris}[1]{{#1}\xspace}
\newcommand{\eqbydef}{\ensuremath{\stackrel{\mathrm{def}}{=}}}
\newcommand{\const}{\ensuremath{\mathrm{const}}}
\newcommand*{\Tren}{\ensuremath{\mT_{ren}}}

\newgray{grayb}{0.5}

\newcommand{\always}{\raisebox{-.2ex}{
			   \mbox{\unitlength=0.9ex
			   \begin{picture}(2,2)
			   \linethickness{0.06ex}
			   \put(0,0){\line(1,0){2}}
			   \put(0,2){\line(1,0){2}}
			   \put(0,0){\line(0,1){2}}
			   \put(2,0){\line(0,1){2}}
			   \end{picture}}}
		      \,}

\renewcommand*{\next}{\!\raisebox{-.2ex}{ %possibly add a little space before
			\mbox{\unitlength=0.9ex%
			\begin{picture}(2,2)%
			\linethickness{0.06ex}%
			\put(1,1){\circle{2}} % Draws circle with
			\end{picture}}}       % diameter 2 at centre 1,1
			\,}

\let\imp=\Rightarrow

\def\ltrue{\hbox{\rm\bf true}}

\newcommand{\dom}[1]{\protect\mathit{dom}(#1)}  % domain of substitution #1
\begin{document}
\title{Monodic Temporal Resolution}
\author{Anatoly Degtyarev\inst{1} 
   \and Michael Fisher\inst{2} 
   \and Boris Konev\inst{2}\thanks{On leave from Steklov Institute of
     Mathematics at St.Petersburg}}  
\institute{%
           Department of Computer Science, King's College, 
           Strand, London WC2R 2LS, U.K.\\
           \email{anatoli@dcs.kcl.ac.uk}
\and      
           Department of Computer Science, University of Liverpool,
           Liverpool L69 7ZF, U.K.\\ 
           \email{\{M.Fisher,$\;$B.Konev\}@csc.liv.ac.uk}}
\date{ }
\maketitle
\begin{abstract}
Until recently, First-Order Temporal Logic (\fotl{}) has been little
understood. While it is well known that the full logic has no finite
axiomatisation, a more detailed analysis of fragments of the logic was
not previously available. However, a breakthrough by Hodkinson et.al.,
identifying a finitely axiomatisable fragment, termed the
\emph{monodic} fragment, has led to improved understanding of
\fotl{}. Yet, in order to utilise these theoretical advances, it is
important to have appropriate proof techniques for the monodic
fragment.

In this paper, we modify and extend the clausal temporal resolution
technique, originally developed for propositional temporal logics, to
enable its use in such monodic fragments. We develop a specific normal
form for formulae in \fotl{}, and provide a complete resolution
calculus for formulae in this form. Not only is this clausal
resolution technique useful as a practical proof technique for certain
monodic classes, but the use of this approach provides us with
increased understanding of the monodic fragment. In particular, we
here show how several features of monodic \fotl{} are established as
corollaries of the completeness result for the clausal temporal
resolution method. These include definitions of new decidable monodic
classes, simplification of existing monodic classes by reductions, and
completeness of clausal temporal resolution in the case of monodic
logics with expanding domains, a case with much significance in both
theory and practice.
\end{abstract}

\section{Introduction}
\label{sec:introduction}
Temporal Logic has achieved a significant role in Computer Science, in
particular, within the formal specification and verification of
concurrent and distributed
systems~\cite{Pnu77,MannaPnueli92:book,Holzmann97e}. While First-Order
Temporal Logic (\fotl{}) is a very powerful and expressive formalism
in which the specification of many algorithms, protocols and
computational systems can be given at the natural level of
abstraction, most of the temporal logics used remain essentially
propositional. The reason for this is that it is easy to show that
\fotl{} is, in general, incomplete (that is, not
recursively-enumerable~\cite{SzaHol88}). In fact, until recently, it
has been difficult to find \emph{any} non-trivial fragment of \fotl{}
that has reasonable properties. A breakthrough by Hodkinson
\emph{et. al.}~\cite{HWZ00} showed that \emph{monodic} fragments of
\fotl{} could be complete, even decidable. (In spite of this, the
addition of equality or function symbols can again lead to the loss of
recursively enumerability from these monodic
fragments~\cite{WZ:APAL:AxMono,DFL02:StudiaLogica}.)

Following the definition of the monodic fragment, work analysing and
extending this fragment has continued rapidly, and holds great promise
for increasing the power of logic-based formal methods. However, until
recently, there were no proof techniques for monodic fragments of
\fotl{}s. Although a tableaux based approach was proposed
in~\cite{KLWZ02}, we here provide a complete resolution calculus for
monodic \fotl{}, based on our work on clausal temporal resolution over
a number of years~\cite{Fisher91,FDP01,DF01,DFK02,DFK03CADE}. The
clausal resolution technique has been shown to be one of the most
effective proof techniques for propositional temporal
logics~\cite{HustadtKonev03:CADE}, and we have every reason to believe
that it will be as least as successful in the case of \fotl{}; this
paper provides the key formal background for this approach.

The structure of the paper is as follows. After a brief introduction
to \fotl{} (Section~\ref{sec:preliminaries}), we define a normal form
that will be used as the basis of the resolution technique and show
that any monodic temporal problem can be transformed into the normal
form (Section~\ref{sec:dsnf}). In Section~\ref{sec:ccalc} we present
the temporal resolution calculus and, in Section~\ref{sec:proof}, we
provide detailed completeness results.

In Sections~\ref{sec:extension} and \ref{sec:properties}, we adapt the
resolution technique to a number of variations of monodic \fotl{},
whose completeness follows from the corresponding adaptation of the
completeness results given in Section~\ref{sec:proof}. Thus, in
Section~\ref{sec:extension}, we provide an extension of the monodic
fragment (as defined in~\cite{HWZ00}) and, in
Section~\ref{sec:properties}, we restrict first-order quantification
in a number of ways to provide sub-classes which admit simplified
clausal resolution techniques.

In the penultimate part of the paper, we examine results relating to
the practical use of the clausal resolution calculus. The first such
aspect concerns decidability, which we consider in
Section~\ref{sec:classes}. An appropriate \emph{loop search} algorithm
is required for implementation of the clausal resolution technique,
and the definition and completeness of such an algorithm is examined
in Section~\ref{sec:loopsearch}. In order to develop a practical
clausal resolution system, as well as examining a fragment with
important applications and a simplified normal form, we present
results relating to resolution over the monodic fragment with
\emph{expanding domains} in Section~\ref{sec:expanding}. This provides
the basis for the system currently being implemented~\cite{KDDFH03}.

Finally, in Section~\ref{sec:concl}, we present conclusions and
outline our future work.

\section{First-Order Temporal Logic}
\label{sec:preliminaries}
First-Order (linear time) Temporal Logic, \fotl, is an extension of
classical first-order logic with operators that deal with a linear and
discrete model of time (isomorphic to $\Nat$, and the most commonly used model
of time). The first-order temporal language is constructed in a standard
way~\cite{Fisher97,HWZ00} from:
%\begin{description}
%\item[predicate symbols] 
\emph{predicate symbols} $P_0, P_1,\dots$ each of which is of some fixed arity
(null-ary predicate symbols are called \emph{propositions});
%\item[individual variables] $x_0, x_1,\dots$;
\emph{individual variables} $x_0, x_1,\dots$;
%\item[individual constants] $c_0,c_1,\dots$;
\emph{individual constants} $c_0,c_1,\dots$;
%\item[booleans] $\land$, $\lnot$, $\lor$, $\implies$, $\equiv$
\emph{Boolean operators} $\land$, $\lnot$, $\lor$, $\implies$, $\equiv$
$\true$ (`true'), $\false$ (`false'); 
\emph{quantifiers} $\forall$ and $\exists$;
together with 
%\item[temporal operators] $\always$ (`always in the future'), 
\emph{temporal operators} $\always$ (`always in the future'),
$\sometime$ (`sometime in the future'), $\next$ (`at the next moment'),
$\until$ (until), and $\unless$ (weak until).
There are no function symbols or equality in this \fotl{} language, but
it does contain constants.
%\end{description}
%
For a given formula, $\phi$, $\const(\phi)$ denotes the set of constants
occurring in $\phi$. We write $\phi(x)$ to indicate that $\phi(x)$ 
has \emph{at most one} free variable $x$ (if not explicitly stated
otherwise). 

Formulae in \fotl are interpreted in \emph{first-order temporal
structures} of the form $\gM = \langle D, I \rangle$, where $D$ is a
non-empty set, the \emph{domain} of $\gM$, and $I$ is a function
associating with every moment of time, $n\in\Nat$, an interpretation
of predicate and constant symbols over $D$.
We require that the interpretation of constants is \emph{rigid}. Thus, for
every constant $c$ and all moments of time $i,j\geq 0$, we have $I_i(c)=
I_j(c)$. The interpretation of predicate symbols is flexible.

A \emph{(variable)
assignment} $\ga$ over $D$ is a function from the set of individual
variables to $D$.
%We require that the interpretation of variables and
%constants is \emph{rigid}, e.g., neither assignments nor
%interpretations of constants depend on time. 
For every moment of time, $n$, there is a corresponding \emph{first-order}
structure $\gM_n = \langle D, I_n\rangle$, where $I_n=I(n)$. Intuitively,
\fotl formulae are interpreted in sequences of \emph{worlds},
$\gM_0,\gM_1,\dots$ with truth values in different worlds being connected by
means of temporal operators.

The \emph{truth} relation $\gM_n\models^\ga \phi$ in a structure
$\gM$, for an assignment $\ga$, is defined inductively in the usual
way under the following understanding of temporal operators:
$$
\begin{array}{lcl}
\gM_n\models^\ga\next\phi & \textrm{iff}& \gM_{n+1}\models^\ga\phi;\\
\gM_n\models^\ga\sometime\phi & \textrm{iff}& \textrm{there exists } m\geq n \textrm{ such that } \gM_{m}\models^\ga\phi; \\
\gM_n\models^\ga\always\phi & \textrm{iff}& \textrm{for all $m\geq n$, } \gM_m\models^\ga\phi;\\
\gM_n\models^\ga(\phi\until\psi) & \textrm{iff}& \textrm{there exists $m\geq n$, such
       that } \gM_m\models^\ga\psi,\\
 & & \hphantom{M}\textrm{and for all $i\in\Nat$, $n\leq i < m$ implies } \gM_m\models^\ga\phi;\\
\gM_n\models^\ga(\phi\unless\psi) & \textrm{iff }& \gM_n\models^\ga(\phi\until\psi)\textrm{ or } \gM_n\models^\ga\always\phi.\\
\end{array}
$$
$\gM$ is a \emph{model} for a formula $\phi$ (or $\phi$ is \emph{true} in
$\gM$) if there exists an assignment $\ga$ such that $\gM_0\models^\ga\phi$. 
A formula is \emph{satisfiable} if it has a model. A formula is \emph{valid}
if it is true in any temporal structure under any assignment.
%
%\paragraph{The monodic fragment.}
%It is known that even ``small'' fragments of \fotl, such as the
%\emph{two-variable monadic} fragment, are not recursively
%enumerable~\cite{Merz:Incomp:1992,HWZ00}. An \fotl-formula $\phi$ is \emph{monodic} if
%any subformulae of the form $\mT\psi$, where $\mT$ is a temporal
%operator, contains at most one free variable. The set of valid monodic
%formulae is known to be finitely axiomatisable~\cite{WZ:APAL:AxMono}.

This logic is complex.
It is known that even ``small'' fragments of \fotl, such
as the \emph{two-variable monadic} fragment (all predicates are unary), are not
recursively enumerable~\cite{Merz:Incomp:1992,HWZ00}. 
However, the set of valid \emph{monodic} formulae  is
known to be finitely axiomatisable~\cite{WZ:APAL:AxMono}.
\begin{definition}[Monodic Formula]
An \fotl-formula $\phi$ is 
called \emph{monodic} if any subformulae of the form $\mT\psi$, where $\mT$ is
one of $\next$, $\always$, $\sometimes$, contains at most one free variable.
\end{definition}
The addition of either equality or function symbols to the monodic fragment 
leads to the loss of recursive enumerability~\cite{WZ:APAL:AxMono}.
Moreover, it was proved in \cite{DFL02:StudiaLogica} that the \emph{two variable
monadic monodic fragment  with equality} is not recursively enumerable.
However, in~\cite{Hodkinson00} it was shown that the \emph{guarded monodic
fragment with equality} is decidable.

\section{Divided Separated Normal Form (DSNF)} 
\label{sec:dsnf}
As in the case of classical resolution, our method works on temporal
formulae transformed into a normal form. The normal form we use
follows the spirit of Separated Normal Form (SNF)
\cite{Fisher91,FDP01} and First-Order Separated Normal Form (SNF$_f$)
\cite{FISHER:CADE92,Fisher97}, but is refined even further.

The development of SNF/SNF$_f$ was partially devised in order to
saparate past, present and future time temporal formula (inspired by
Gabbay's separation result \cite{GABBAY:1987}). Thus, formulae in
SNF/SNF$_f$ comprise implications with present-time formulae on the
left-hand side and (present or) future formulae on the right-hand
side. The transformation of temporal formulae into separated form is
based upon the well-known \emph{renaming}
technique~\cite{Tseitin68,PlGr:Renaming:86}, which preserves
satisfiability and admits the extension to temporal logic in (Renaming
Theorems \cite{Fisher97}).

Another aim with SNF/SNF$_f$ was to reduce the variety of temporal
operators used to a simple core set. To this end, the transformation
to SNF/SNF$_f$ involves the removal of temporal operators represented
as \emph{maximal} fixpoints, that is, $\always$ and $\unless$ (Maximal
Fixpoint Removal Theorems \cite{Fisher97}). Note that the $\until$
operator can be represented as a combination of operators based upon
maximal fixpoints and the $\sometime$ operator (which is retained
within SNF/SNF$_f$). This transformation is based upon the simulation
of fixpoints using QPTL \cite{Wolper:PhD:82}.

In the first-order context, we now add one further aim, namely to
divide the temporal part of a formula and its (classical) first-order
part in such way that the temporal part is as simple as possible. The
modified normal form is called Divided Separated Normal Form
%\footnote{A more appropriate name
%for this refinement might be ``Even More Separated Normal Form'' by
%analogy with \cite{BHSch:93:Lib}},
or DSNF for short.

\begin{definition}[Temporal Step Clauses]
A \emph{temporal step clause} is a formula either of the form
$l\imp\next m$, where $l$ and $m$ are propositional
literals, or $(L(x)\imp\next M(x))$, where $L(x)$ and $M(x)$
are unary literals.
We call a clause of the the first type an
(original) \emph{ground} step clause, and of the second type an
(original) \emph{non-ground} step clause\footnote{We could also allow
arbitrary Boolean combinations of propositional and unary literals in 
the right hand side of ground and non-ground step clauses, respectively, and
all results of this paper would hold. We restrict ourselves with literals for
simplicity of the presentation.}.
\end{definition}
\begin{definition}[DSNF]
A \emph{monodic temporal problem  in Divided
Separated Normal Form (DSNF)} is a quadruple 
$\langle \uP, \iP, \sP, \eP\rangle$, where
\begin{enumerate}
\item the universal part, $\uP$, is a finite set of arbitrary closed first-order
      formulae;% (clauses);
\item the initial part, $\iP$, is, again, a finite set of arbitrary closed first-order
      formulae; 
\item the step part, $\sP$, is a finite set of original (ground and non-ground) 
      temporal step clauses; and
\item the eventuality part, $\eP$, is a finite set of 
eventuality clauses of the form $\sometime L(x)$ (a \emph{non-ground}
eventuality clause) and $\sometime l$ (a \emph{ground eventuality}
clause), where $l$ is a propositional literal and $L(x)$ is a unary
non-ground literal.
\end{enumerate}
%% The sets $\uP$, $\iP$, $\sP$, and $\sP$ are finite.
\end{definition}
Note that, in a monodic temporal problem, we disallow two
different temporal step clauses with the same left-hand 
sides. This requirement can be easily guaranteed by renaming.
%Any  problem with
%the same left-hand sides can be easily transformed by renaming into one 
%without.
%%%
% \noindent A literal $l$ from an eventuality clause is called an
% \emph{eventuality literal}.  Step clauses will also be referred to as
% \emph{step rules}.  Without loss of generality, we can assume that
% there are no two different temporal step clauses with the same
% left-hand sides.
% 
% 

In what follows, we will not distinguish between a finite set of
formulae ${\mathcal X }$ and the conjunction $\bigwedge {\mathcal X}$
of formulae within the set. With each  monodic temporal
problem, we associate the formula
$$
\iP\land\always\uP\land\always\forall x\sP\land\always\forall x\eP.
$$
Now, when we talk about particular properties of a temporal problem (e.g.,
satisfiability, validity, logical consequences etc) we mean properties of
the associated formula.

%\subsection{Translation into DSNF}

Arbitrary monodic first-order temporal formula can be transformed into a 
satisfiability-equivalent unconditional eventuality monodic temporal
problem. We present the transformation as a two stage reduction.
\paragraph{Reduction to conditional DSNF.}
We first give a reduction from monodic FOTL to a normal form where, in addition
to the parts above, \emph{conditional} eventuality clauses of the form
$$
P(x)\implies\sometime L(x)  \textrm{ and }
p\implies\sometime l
$$
are allowed. The reduction is based on using a renaming technique to substitute
non-atomic subformulae and replacing temporal operators by their fixed point
definitions described e.g. in~\cite{FDP01}. The translation can be described
as a number of steps.
\begin{enumerate}
\item Translate a given monodic formula to negation normal form.
(To assist understanding of the translation, we list here some equivalent FOTL
formulae.)
$$
\begin{array}{lcl}
\forall x (\lnot\next\phi(x)&\equiv&\next\lnot\phi(x)); \\
\forall x (\lnot\always\phi(x)&\equiv&\sometime\lnot\phi(x));\\
\forall x (\lnot\sometime\phi(x)&\equiv&\always\lnot\phi(x)\\
\forall x (\lnot(\phi(x)\until\psi(x))&\equiv&\lnot\psi(x)\unless(\lnot\phi(x)\land\psi(x))));\\
\forall x (\lnot(\phi(x)\unless\psi(x))&\equiv&\lnot\psi(x)\until(\lnot\phi(x)\land\psi(x))).
\end{array}
$$
\item Recursively rename innermost temporal subformulae, $\next\phi(x)$, 
$\sometime\phi(x)$, $\always\phi(x)$, $\phi(x)\until\psi(x)$,
$\phi(x)\unless\psi(x)$ by a new unary predicate $P(x)$. Renaming introduces
formulae defining $P(x)$ of the following form:
$$
\begin{array}{lcl}
   (a)\;\always \forall x (P(x)&\imp& \next \phi(x));\\
   (b)\;\always \forall x (P(x)&\imp& \sometime \phi(x));\\
   (c)\;\always \forall x (P(x)&\imp& \always \phi(x)); \\
   (d)\;\always \forall x (P(x)&\imp& \phi(x)\until\psi(x))\\
   (e)\;\always \forall x (P(x)&\imp& \phi(x)\unless\psi(x)).
\end{array}
$$
Formulae of the form $(a)$ and $(b)$ are in the normal form\footnote{Possibly, 
after (first-order) renaming the complex expression $\phi(x)$; the formulae
introduced by renaming  are put in the universal part. This kind of first-order
renaming is used implicitly further in this section.}, formulae of the form
$(c)$ and $(d)$ require extra
reduction by
removing the temporal operators using their fixed point definitions;
formulae of the last kind can be reduced by the semantics of the $\unless$
operator.
\item Use fixed point definitions\\
$\always\forall x (P(x)\imp\always \phi(x))$
is satisfiability equivalent to
$$
\begin{array}{l}
\always \forall x (P(x)\implies R(x))\\
\land \always \forall x (R(x)\implies\next R(x))\\
\land \always \forall x (R(x)\implies \phi(x)),
\end{array}
$$
and $\always\forall x (P(x)\implies(\phi(x)\until\psi(x)))$
is equivalent (w.r.t. satisfiability) to
$$
\begin{array}{l}
\always \forall x (P(x)\implies \sometime\psi(x))\\
\land \always \forall x (P(x)\implies \phi(x)\lor\psi(x))\\
\land \always \forall x (P(x)\implies S(x)\lor\psi(x))\\
\land \always \forall x (S(x)\implies \next(\phi(x)\lor\psi(x)))\\
\land \always \forall x (S(x)\implies \next(S(x)\lor\psi(x))),
\end{array}
$$
where $R(x)$ and $S(x)$ are new unary predicates.%\\[1mm]
%where $S(x)$ is a new unary predicate.
\end{enumerate}
\paragraph{Conditional problems to unconditional problems.}\mbox{}
In the second stage, we replace any formula 
$\always\forall x (P(x)\implies\sometime L(x))$ 
by 
\begin{eqnarray}
& &\always\forall x ( ((P(x) \andl \lnot L(x)) \implies \wfL(x)))\label{equ:rm1}\\ 
& &\always\forall x ((\wfL(x) \andl \next \lnot L(x) ) \implies \next \wfL(x))\label{equ:rm3}\\
& &\always\forall x (\sometime \lnot \wfL(x))\label{equ:rm2}
\end{eqnarray}%stopzone
where $\wfL(x)$ is a new unary predicate. 
\begin{lemma}\label{th:unconditioning}
  \quad
  $\Phi \cup \{ \always\forall x ( P(x) \implies  
  \sometime L(x) ) \}$\, is satisfiable if, and only if,\, 
   $\Phi\cup \{(\ref{equ:rm1}),(\ref{equ:rm3}),(\ref{equ:rm2})\}$ is
   satisfiable.
\end{lemma}
\begin{proof}
($\Rightarrow$) \quad Let $\gM$ be a model of 
$\Phi \cup \{ \always \forall x ( P(x) \implies  \sometime L(x) ) \}$. 
Let us extend this model by a new
predicate $\wfL$ such that, in the extended model, $\gM'$, 
formulae (\ref{equ:rm1}), (\ref{equ:rm3}), and (\ref{equ:rm2})
would be true. 

Let $d$ be an arbitrary element of the domain $D$.
We define the truth 
value of $\wfL(d)$ in $n$-th moment, $n\in\Nat$,
depending on whether
$\gM \models \always \sometime P(d)$ or  
$\gM \models \sometime \always \notl P(d)$.

\begin{itemize}
\item Assume  $\gM \models \always \sometime P(d)$. 
Together with $\gM\models\always\forall x ( P(x) \implies \sometime L(x) )$, this implies that
  $\gM \models \always \sometime L(d)$.

For every 
$n \in\Nat $ let us put
$$
\gM'_n \models  \lnot \wfL(d)\, \Leftrightarrow\, \gM'_n \models  L(d)\,
\qquad( \Leftrightarrow\, \gM_n \models  L(d)).
$$
\item Assume $\gM \models \sometime \always \notl P(d)$. 
There are two possibilities: 
\begin{itemize}
\item $\gM \models \always \notl P(d)$. In this case let us put
$\gM'_n \models \lnot \wfL(d)$ for all $n \in\Nat $.
\item There exists $m\in\Nat$ such that 
$ \gM_m \models P(d)$ and, for all $n > m$, $\gM_n \models \notl
 P(d)$. These conditions imply, in particular, that there is $l \geq m$
 such that $\gM_l \models L(d)$ if the formula is satisfiable. Now we
 define $\wfL(d)$ in $\gM'$ as follows:
$$
\begin{array}{lll}
\gM'_n \models  \lnot \wfL(d)\, \Leftrightarrow\, \gM'_n \models  L(d)
& \mbox{ if } & 0\leq n < l, \\
\gM'_n \models \lnot \wfL(d) & \mbox{ if } &  n \geq l. \\ 
\end{array}
$$
\end{itemize}

\end{itemize}
It is easy to see that $\gM'$ is the required model.
\smallskip

($\Leftarrow$)\quad Let us show that $\always\forall x ( P(x) \implies  
  \sometime L(x) )$ is a logical consequence of
  $ \Phi\cup\{(\ref{equ:rm1}),(\ref{equ:rm3}), (\ref{equ:rm2})\}$. 

\noindent Let $\gM'$ be a model of $ \Phi\cup\{(\ref{equ:rm1}),(\ref{equ:rm3}),(\ref{equ:rm2})\}$. By contradiction, suppose
$\gM' \not\models \always\forall x ( P(x) \implies  \sometime L(x) )$, 
that is,  $\gM' \models \sometime \exists x ( P(x) \andl  
\always \lnot L(x) )$. Let  $m \in {\mathbb N}$ be an index and $e\in D_m$
be a domain element
such that $\gM'_m \models P(e)$ and for all $n \geq m$, $\gM'_n
\models \lnot L(e) )$.  Then from (\ref{equ:rm1}) and (\ref{equ:rm3}) we
conclude that for all $n \geq m$, we have $\gM'_n \models \wfL(e)
)$. However, this conclusion contradicts the formula $\always\forall x
\sometime \lnot \wfL(x)$ which is true in $\gM'$.
\end{proof}  
This concludes to the following theorem.
\begin{theorem}[Transformation]
Every monodic first-order temporal formula can be reduced, in a
satisfiability equivalence preserving way, to DSNF with
at most a linear increase in size of the problem.
\end{theorem}

% \begin{example}\label{ex:dsnf1}
% Let us consider the temporal formula
% $\exists x \always \forall y \forall z \exists u \Phi(x,y,z,u)$, where
% $\Phi(x,y,z,u)$ does not contain temporal
% operators, and reduce it to DSNF.
%
% First, we rename the temporal subrormula by a new predicate,
% $$
% \exists x P_1(x) \land \always\forall x
%    (P_1(x)\implies \always \forall y \forall z\exists u \Phi(x,y,z,u)).
% $$
% Then we ``unroll'' the \emph{always} operator
% $$
% \exists x P_1(x) \land \always\forall x(P_1(x)\implies\next P_1(x))\land
%   \always\forall x ( P_1(x)\implies \forall y \forall z\exists u \Phi(x,y,z,u)).
% $$
% The parts of this formula form the following monodic temporal problem:
% $\iP=\{\exists x P_1(x)\}$,
% $\uP=\{\forall x (P_1(x)\implies \forall y\forall z \exists u \Phi(x,y,z,u))\}$,
% $\sP=\{ \forall x P_1(x)\implies\next P_1(x) \}$,
% $\eP=\emptyset$.
% \end{example}
\begin{example}\label{ex:dsnf2}
Let us consider the temporal formula $\exists x \always \sometime
\forall y \forall z \exists u \,\Phi(x,y,z,u)$ where $\Phi(x,y,z,u)$
does not contain temporal operators and reduce it to DSNF. First,
we rename the inmost temporal subformula by a new predicate,
%As in the previous
%case, we reduce it to
$$
 \exists x \always P_1(x) \land
 \always\forall x[P_1(x)\implies \sometime \forall y \forall z\exists
 u \,\Phi(x,y,z,u)].
$$
Now, we rename the first `$\always$'-formula and the subformula
under the `$\sometime$' operator,
 \begin{eqnarray*}
\exists x P_3(x) &\land&
 \always\forall x[P_1(x)\!\implies\!\sometime P_2(x)]\\
 & \land &
 \always\forall x [P_2(x)\!\implies\!\forall y \forall z\exists u
 \,\Phi(x,y,z,u)]
\\  &  \land &
\always\forall x [P_3(x)\!\implies\!\always P_1(x)],
\end{eqnarray*}
``unwind'' the `$\always$' operator
\begin{eqnarray*}
\exists x P_3(x) &\land & \always\forall x[P_1(x)\implies \sometime P_2(x)]\\
&\land& \always\forall x [P_2(x)\implies \forall y \forall z\exists u \,\Phi(x,y,z,u)]\\
 &  \land &  \always\forall x [P_3(x)\implies P_4(x)]\\
 &\land& \always\forall x [P_4(x)\implies \next P_4(x)]\\
 & \land & \always\forall x [P_4(x)\implies P_1(x)],
\end{eqnarray*}
and, finally, reduce the conditional eventuality to an unconditional one. 
$$
\begin{array}{lcl}
\exists x P_3(x) &\land & \always\forall x [P_2(x)\implies \forall y \forall
z\exists u \,\Phi(x,y,z,u)]\\
  & \land &   \always\forall x [P_3(x)\implies P_4(x)]\\
 & \land & \always\forall x [P_4(x)\implies \next P_4(x)]\\
 & \land & \always\forall x [P_4(x)\implies P_1(x)]\\
&\land & \always\forall x [(P_1(x)\land\lnot P_2(x))\implies \mathit{waitforP_2}(x)] \\
&\land & \always\forall x [(\mathit{waitforP_2}(x)\land\next\lnot P_2(x))\implies
      \next \mathit{waitforP_2}(x)] \\
&\land & \always\forall x \sometime \lnot \mathit{waitforP_2}(x).
\end{array}
$$
The parts of this formula form the following monodic temporal problem (we
also rename the complex $P_2(x)\lor\mathit{waitforP_2}(x)$
expression by $P_5(x)$):
\smallskip

$$
\begin{array}{lcl}
\iP&=&\left\{\begin{array}{l}
  \exists x P_3(x)
\end{array}\right\},\\[2mm]
\uP&=&\left\{
\begin{array}{l}
  \forall x (P_2(x)\implies \forall y \forall z\exists u \,\Phi(x,y,z,u)),\\
  \forall x (P_3(x)\implies P_4(x)),\\
  \forall x (P_4(x)\implies P_1(x)),\\
  \forall x ((P_1(x)\land\lnot P_2(x))\implies \mathit{waitforP_2}(x)),\\
  \forall x (P_5(x)\implies P_2(x)\lor\mathit{waitforP_2}(x))
\end{array}\right\},\\[10mm]
\sP&=&\left\{
\begin{array}{l}
  P_4(x)\implies\next P_4(x),\\
  \mathit{waitforP_2}(x) \implies \next P_5(x) 
\end{array}\right\}\\[4mm]
\eP&=&\left\{\begin{array}{l}
 \always\forall x \sometime\lnot \mathit{waitforP_2}(x)
\end{array}\right\}. 
\end{array}
$$
\hfill\qed
\end{example}

\section{Temporal resolution}
\label{sec:ccalc}
As in the propositional case~\cite{Fisher91,DFK02}, our calculus works
with \emph{merged step clauses}, but here the notion of a merged step
clauses is much more complex. This is, of course, because of the
first-order nature of the problem and the fact that skolemisation is
not allowed under temporal operators. In order to build towards the
calculus, we first provide some important definitions.

\begin{definition}[\Grounded{} Step Clauses]
Let $\TProb$ be a monodic temporal problem, and let
\begin{equation}\label{eq:premises}
P_{i_1}(x)\implies\next M_{i_1}(x), \dots, P_{i_k}(x)\implies\next M_{i_k}(x)
\end{equation}
be a subset of the set of its original non-ground step clauses. Then
\begin{eqnarray}
\forall x (P_{i_1}(x)\lor\dots\lor P_{i_k}(x))&\implies&\next\forall x (M_{i_1}(x)\lor\dots\lor M_{i_k}(x)),\label{eq:forall}\\
\exists x (P_{i_1}(x)\land\dots\land P_{i_k}(x))&\implies&\next\exists x (M_{i_1}(x)\land\dots\land M_{i_k}(x)),\label{eq:exists}\\
P_{i_j}(c)&\implies&\next M_{i_j}(c)\label{eq:const}
\end{eqnarray}
are \emph{\grounded} step clauses, where $c\in\const(\TProb)$ and $j = 1\dots
k$.   
\end{definition}
A \grounded{} step clause is a logical consequence of its premises obtained by
``dividing'' and bounding left-hand and right-hand sides.

\begin{definition}[Merged \Grounded{} Step Clauses]
Let $\{\Phi_1\implies\next\Psi_1, \dots, \Phi_n\implies\next\Psi_n\}$
be a set of \grounded{} step clauses or original \emph{ground} step clauses. Then
$$
\bigwedge\limits_{i=1}^n \Phi_i \imp \next
\bigwedge\limits_{i=1}^n {\Psi_i}
$$
is called a \emph{merged \grounded{} step clause}.  
\end{definition}
Note that the
left-hand and right-hand sides of any merged \grounded{} step clause are
closed formulae.
\begin{definition}[Full Merged Step Clauses] 
Let
$
\mA\implies\next\mB
$
be a merged \grounded{} step clause,
$
P_{1}(x)\implies \next M_{1}(x), \dots, P_{k}(x)\implies \next M_{k}(x)
$
be original step clauses, and 
$
A(x)\eqbydef\bigwedge\limits_{i=1}^k P_i(x),
$ $%\quad
B(x)\eqbydef\bigwedge\limits_{i=1}^k M_i(x).
$
Then
$$
\forall x(\mA\land A(x)\implies\next(\mB\land B(x)))
$$ 
is called a \emph{full merged step clause}\label{page:merged}. In the
case $k=0$, the conjunctions $A(x)$, $B(x)$ are empty, that is,
their truth value is $\true$, and
the merged step clause is just a merged \grounded{} step clause.
\end{definition}
\begin{definition}[Constant Flooding]
Let $\TProb$ be a monodic temporal problem, 
$
\TProb^c = \TProb\cup\{\sometime L(c)\;|\; \sometime L(x)\in\eP, c\in
\const(\TProb) \}
$ 
is the \emph{constant flooded form}\footnote{\moreboris{Strictly speaking, $\TProb^c$ is not in DSNF: we have to rename ground
eventualities by propositions. Rather than ``flooding'', we could have introduced special inference rules to deal
with constants.}} of $\TProb$.
\end{definition}
Evidently, $\TProb^c$ is satisfiability equivalent to $\TProb$.
%
%Let $\TProb$ be a monodic temporal problem, and let $\gC$ be the
%\moreboris{set of constants occurring in $\TProb$}. Now, denote by $\TProb^c$
%the set $\TProb\cup\{\sometime L(c)\;|\; \sometime L(x)\in\eP, c\in\gC \}$. If
%$\TProb = \TProb^c$, then we say that the temporal problem is
%\emph{constant flooded} 
%% (Strictly speaking, these ground
%% eventualities extend our definition of the normal form; however,
%% they may be renamed by propositions.)
%
\begin{example}\label{ex:res2-1}
Let us consider a temporal problem given by
$$
\begin{array}{lcl}
\iP &=& \left\{
\begin{array}{lll}
i1. & & Q(c)\\
\end{array}
\right\},\\[1ex]
\uP &=& \left\{
\begin{array}{lll}
u1. & & \always \exists x (P_1(x)\land P_2(x))\\
u2. & & \always \forall x (Q(x) \andl \exists y (\notl P_1(y)\land \notl P_2(y))
\implies L(x)) \\
\end{array}
\right\},\\[2ex]
\sP &=& \left\{
\begin{array}{llrcl}
s1. & & P_1(x) &\imp& \next \notl P_1(x) \\
s2. & & P_2(x) &\imp& \next \notl P_2(x) \\
s3. & & Q(x) &\imp& \next Q(x)\\
\end{array}
\right\}\\[4ex]
\eP &=& \left\{
\begin{array}{lll}
e1. & & \sometime \notl L(x)\\
\end{array}
\right\},\;
\end{array}
$$
Then 
$$
\begin{array}{lcrcl}
d1. & & P_1(c) & \imp & \next \notl P_1(c), \\
d2. & & \exists y P_1(y) & \imp & \next \exists y \notl P_1(y), \\
d3. & & \forall y P_1(y) & \imp & \next \forall y \notl P_1(y), \\
d4. & & \exists y (P_1(y)\land P_2(y))& \imp &\next\exists y(\lnot P_1(y)\land
\lnot P_2(y))\\
d5. & & \forall y (P_1(y)\lor P_2(y))& \imp &\next\forall y(\lnot P_1(y)\lor
\lnot P_2(y))
\end{array}
$$
are examples of derived step clauses. Every derived step clause is also a 
merged derived step clause. In addition, 
$$
\begin{array}{lcrcl}
m1. & & P_1(c)\land \exists y P_1(y)& \imp & \next(\notl P_1(c)\land \exists y \notl P_1(y)),\\
m2. & & \exists y P_1(y)\land \forall y P_1(y) & \imp & \next(\exists y \notl P_1(y)\land \forall y \notl P_1(y))
\end{array}
$$
are other examples of merged derived step clauses. Finally, 
$$
\begin{array}{lcrcl}
fm1. & & \forall x (P_2(x)\land P_1(c)& \imp & \next(\lnot P_2(x)\land\lnot P_1(c))),\\
fm2. & & \forall x (Q(x)\land \exists y (P_1(y)\land P_2(y)) & \imp &\next(Q(x)\land \exists y(\lnot P_1(y)\land
\lnot P_2(y)))  ),\\
fm3. & & \forall x (P_1(x)\land \exists y P_1(y)\land \forall y P_1(y) & \imp & \next(Q(x)\land \exists y \notl P_1(y)\land \forall y \notl P_1(y)))
\end{array}
$$
are examples of full merged step clauses.

Note that, constant flooding adds to the problem the eventuality
$\sometime\lnot L(c)$. 
%In addition to $e1$, the eventuality part of the constant flooded form of the
%problem contains $\sometime \lnot L(c)$.
\end{example}

\subsubsection*{Inference Rules.}
The inference system we use consists of the following inference
rules. (Recall that the premises and conclusion of these rules are
(implicitly) closed under the $\always$ operator.)
%
%\noindent\textbf{Inference Rules. }

In what follows, $\mA \imp \next \mB$ and
$\mA_i \imp \next \mB_i$ denote merged \grounded{} step clauses, $\forall x (\mA \land A(x) \imp \next (\mB \land
B(x)))$ and $\forall x (\mA_i \land A_i(x) \imp \next (\mB_i \land
B_i(x)))$ denote full merged step clauses, and
$\mU$ denotes the (current) universal part of the problem.
\begin{itemize}
\item \emph{Step resolution rule w.r.t. $\mU$}:\quad
$\begin{array}{c}
\infer[(\next_{\mathit{res}}^{\mU})\,,]{\notl \mA}
{ \mA \imp \next \mB }
\end{array}$ where $\uP \cup \{\mB\} \models \perp$.
\item \emph{Initial termination rule w.r.t. $\mU$:}\quad
The contradiction $\perp$ is derived and the derivation is (successfully)
terminated if\, $\uP \cup \iP \models \perp$.
\item \emph{Eventuality resolution rule w.r.t. $\mU$:}
$$
\infer[(\sometime_{\mathit{res}}^{\mU})\,,]{\forall x\bigwedge\limits_{i=1}^n (\notl \mA_i\lor \lnot A_i(x))}
{
\begin{array}{c}
\forall x (\mA_1 \land A_1(x)  \imp  \next (\mB_1 \land B_1(x)))  \\
\dots \\
\forall x (\mA_n \land A_n(x)  \imp  \next (\mB_n \land B_n(x)))  \\
\end{array} &
\quad\sometime L(x)
}
$$
where $\forall x (\mA_i \land A_i(x)  \imp  \next \mB_i \land B_i(x))$
are full merged step clauses such that for all $i \in \{1,\ldots,
n\}$, the \emph{loop} side conditions

$$
\forall x (\uP \land  \mB_i \land B_i(x) \implies \lnot L(x)) \quad  \mbox{and}
\quad
\forall x (\uP  \land \mB_i \land B_i(x) \implies \bigvee\limits_{j=1}^n (\mA_j\land A_j(x))
$$

\noindent are both valid. 

The set of merged step clauses, satisfying the loop side conditions,
is called a \emph{loop in $\sometime L(x)$}\label{page:loop} and the
formula $\bigvee\limits_{j=1}^{n}(\mA_j(x)\land A_j(x))$ is called a
\emph{loop formula}.
\item \emph{Eventuality termination rule w.r.t. $\mU$:} The
contradiction $\perp$ is derived and the derivation is (successfully)
terminated if\, $\uP \models \forall x \lnot L(x)$, where $\sometime
L(x)\in\eP$\footnote{In the case $\uP\models\forall x \lnot L(x)$, the
\emph{degenerate clause}, $\true\implies\next\true$, can be considered
as a premise of the eventuality resolution rule; the conclusion of the
rule is then $\lnot\true$ and the derivation successfully
terminates.}.

\item \emph{Ground eventuality resolution rule w.r.t. $\mU$:}\
$$
\infer[(\sometime_{\mathit{res}}^{\mU})\,,]{(\bigwedge\limits_{i=1}^n \notl \mA_i)}
{ \mA_1  \imp  \next \mB_1, &
\ldots, &
 \mA_n  \imp  \next \mB_n &
\quad\sometime l}
$$
%where $\mA_i \imp \next \mB_i$ are divided merged step clauses such
where $\mA_i \imp \next \mB_i$ are merged grounded step clauses such
that the \emph{loop} side conditions
$$
\uP \land \mB_i  \models \lnot l \quad  \mbox{and} \quad
  \uP  \land \mB_i  \models \bigvee\limits_{j=1}^n  \mA_j
 \quad  \mbox{for all} \quad
      i \in \{1,\ldots, n\}
$$
are satisfied. \emph{Ground loop} and \emph{ground loop formula} are 
defined similarly to the case above.
\item \emph{Ground eventuality termination rule w.r.t. $\mU$:}\\
The contradiction $\perp$ is derived and the derivation is (successfully)
terminated if\, $\uP \models \lnot l$, where $\sometime l\in\eP$ .
\end{itemize}
\begin{definition}[Derivation]
A \emph{derivation} 
%in the described inference systems 
is a sequence of universal parts, $\uP = \uP_0 \subseteq \uP_1 \subseteq \uP_2
\subseteq\dots$, extended little by little by the conclusions of the inference
rules.  Successful termination means that the given problem is unsatisfiable.
The $\iP$, $\sP$ and $\eP$ parts of the temporal problem are not changed in a
derivation.
\end{definition}
%\CADEout{
%\noindent Successful termination means that the given problem is unsatisfiable.
%\begin{note}}
%The conclusion of all our inference rules is universal.  
%\textit{The proof procedure} does not change the $\iP$, $\sP$ or $\eP$ parts of
%the temporal problem; the conclusion of an inference rule is added to 
%$\uP$ which is extended step by
%step.\looseness=-1
%\CADEout{\end{note}}

\begin{note}
The \emph{eventuality resolution rule} above can be thought of as two separate
rules: an induction rule to extract a formula of the form $\forall
x(P(x)\imp\next\always\lnot L(x))$ and a resolution rule to resolve
this with $\forall y\sometime L(y)$, that is,
\begin{itemize}
\item \emph{Induction rule w.r.t. $\mU$:}     
$$
\infer[(ind^{\mU})\,,]{\forall x(\bigvee\limits_{i=1}^n (\mA_i\land A_i(x))
 \implies \next\always\lnot L(x))}
{
\begin{array}{c}
\forall x (\mA_1 \land A_1(x)  \imp  \next (\mB_1 \land B_1(x)))  \\
\dots \\
\forall x (\mA_n \land A_n(x)  \imp  \next (\mB_n \land B_n(x)))  \\
\end{array} &
}
$$
% $$
% \infer[(ind^{\mU})\,,]{(\bigvee\limits_{i=1}^n  A_i) \imp \next \always \notl l }
% { A_1  \imp  \next B_1, & \ldots, & A_n  \imp  \next B_n }
% $$
(with the same side conditions as the eventuality resolution rule above).

The formula $\bigvee\limits_{i=1}^n (\mB_i\land B_i(x))$ can be considered 
as an \emph{invariant formula} since, within the loop detected, this
formula is always true.
\item \emph{Pure eventuality resolution:}
%% \footnote{We could as well formulate this rule in a more
%% ``traditional'' form, with $\next\sometime L(x)$  as the second
%% premise of the rule. Examination of the possible models shows that
%% these two rules are equivalent.} 
$$
\infer[(\sometime_{res})\,.]{\forall x\bigwedge\limits_{i=1}^n (\notl \mA_i\lor \lnot A_i(x))}
{\forall x(\bigvee\limits_{i=1}^n (\mA_i\land A_i(x))
 \implies \next\always\lnot L(x)) &
 \quad \sometime L(x)
}
$$
\end{itemize}
The \emph{ground eventuality resolution rule} can be split into two parts
in a similar way.
\end{note}

%

%\section{Examples}
\begin{example}[Example~\ref{ex:res2-1} contd.]
We apply temporal resolution to the (unsatisfiable) temporal problem from
Example~\ref{ex:res2-1}.
%\noindent \underline{Solution}
%First, we produce the following \grounded{} step clause from 
%$s1$ and $s2$:
%$$
%\begin{array}{lll}
%d1. & & \exists y (P_1(y)\land P_2(y))\implies\next\exists y(\lnot P_1(y)\land
%\lnot P_2(y)).
%\end{array}
%$$
%Then merge $d1$ and $s3$ to give
%$$
%\begin{array}{lll}
%m1. & & \forall x (\exists y( P_1(y)\land P_2(y)) \andl Q(x) \imp \next
%(\exists y (\notl P_1(y)\land\lnot P_2(y) )\andl Q(x))).
%\end{array}
%$$
It can be immediately checked that the loop side conditions are valid for the
full merged step clause $fm2$, 
$$
\begin{array}{lcrcl}
fm2. & & \forall x (Q(x)\land \exists y (P_1(y)\land P_2(y)) & \imp &\next(Q(x)\land \exists y(\lnot P_1(y)\land
\lnot P_2(y)))  ),
\end{array}
$$
that is,
$$
\begin{array}{lll}
\exists y (\notl P_1(y)\land \notl P_2(y)) \andl Q(x) \implies L(x) & \quad & (\mbox{see } u2),\\
\exists y ( \notl P_1(y)\land \notl P_2(y))\andl Q(x) \implies \exists y (P_1(y)\land P_2(y))  \andl Q(x)
& &
(\mbox{see } u1).
\end{array}
$$
We apply the eventuality resolution rule to $e1$ and $m1$ and derive a new
universal clause
$$
\begin{array}{lll}
nu1 . & &  \always \forall x ( \notl (\exists y (P_1(y)\land P_2(y))) \lor \lnot Q(x))
\end{array}
$$
which contradicts clauses $u1$ and $i1$ (the initial termination rule is applied).
\end{example}

\begin{example}
The need for constant flooding can be demonstrated by the following example.
None of the  rules of temporal resolution can be applied directly to the
(unsatisfiable) temporal problem given by 
$$
\begin{array}{ll}
\iP=\{P(c)\}, \qquad & \sP=\{q\imp\next q\},\\
\uP=\{q\equiv P(c)\},\qquad  & \eP=\{\sometime\lnot P(x)\}.
\end{array}
$$
If, however, we add to the problem an eventuality clause $\sometime l$ and a 
universal clause $l\implies\lnot P(c)$, the step clause
$
q\imp\next q
$
will be a loop in $\sometime l$, and the eventuality resolution rule would
derive $\lnot \true$\footnote{Note that the non-ground eventuality
$\sometime\lnot P(x)$ is not used. We show in Section~\ref{sec:properties} that
if all step clauses are ground, for constant flooded problems we can neglect
non-ground eventualities.}.  
\end{example}
Correctness of the presented calculi is straightforward.
\begin{theorem}[Soundness of Temporal Resolution]
The rules of temporal resolution preserve satisfiability.
\end{theorem}
\begin{proof}
Considering models for \fotl{} formulae, it can be shown that the
temporal resolution rules preserve satisfiability. Let $\gM = \langle
D, I\rangle$ be a temporal structure and $\ga$ be a variable
assignment. We assume that a temporal problem $\TProb$ is true in
$\gM$ under the assignment $\ga$ and show that $\TProb$, extended with
the conclusion of a temporal resolution rule, is true in $\gM$ under
$\ga$. We do this by considering cases of the inference rule used, as
follows.
\begin{itemize}
\item Consider the step resolution rule. Let $\mA\implies\next\mB$ be
  a merged derived clause and assume that $\gM_0\models^\ga
  \always(\mA\implies\next\mB)$, $\uP\cup\mB\models\perp$, but for
  some $i\geq0$, $\gM_i\not\models^\ga \lnot\mA$. Then
  $\gM_{i+1}\models^\ga\mB$ in contradiction with the side condition
  of the rule.

\item Consider now the eventuality resolution rule. 
Let $\forall x (\mA_i \land A_i(x)  \imp  \next \mB_i \land B_i(x))$,
$i \in \{1,\ldots,n\}$, be 
full merged step clauses and $\sometime L(x)$ be an eventuality
such that 
$\gM_0\models^\ga \bigwedge\limits_{i=1}^n \forall x (\mA_i \land A_i(x)  \imp  \next \mB_i \land
B_i(x))$,
$\gM_0\models^\ga \always\forall x \sometime L(x)$,
and the loop side conditions
$
\forall x (\uP \land  \mB_i \land B_i(x) \implies \lnot L(x))$ and
$\forall x (\uP  \land \mB_i \land B_i(x) \implies \bigvee\limits_{j=1}^n (\mA_j\land A_j(x))
$
are both valid, but for some $k\geq0$, $\gM_k\not\models^\ga \forall
x\bigwedge\limits_{i=1}^n (\notl \mA_i\lor \lnot A_i(x))$. It follows 
there exists a domain element $d\in D$ such that 
$\gM_k\models^\ga(\mA_j \land  A_j(d))$. It is not hard to see that,
by validity of the loop side conditions and by the fact that the full merged 
clauses are true in $\gM$ under $\ga$,
$\gM_l\models^\ga\lnot L(d)$ for all $l>k$, that is,
$\gM_{k+1}\models^\ga\always\lnot L(d)$ in contradiction with the
eventuality. 

\item Correctness of the initial termination and eventuality
  termination rules is obvious.

\item Correctness of the ground counterparts of the eventuality
  resolution and eventuality termination rules can be proved in a
  similar way.
\end{itemize}
\end{proof}
%%%%%
We formulate now the completeness result and prove it in 
Section~\ref{sec:proof}, which is entirely devoted to this issue.
\begin{theorem}[Completness of Temporal Resolution]\label{th:complete}%
Let an arbitrary monodic temporal problem $\TProb$ be
unsatisfiable.  Then there exists a successfully
terminating derivation by temporal resolution from $\TProb^c$.  
\end{theorem}

\section{Completeness of Temporal Resolution}
\label{sec:proof}
In short, the proof of Theorem~\ref{th:complete} proceeds by building
a graph associated with a monodic temporal problem, then showing that
there is a correspondence between properties of the graph and of the
problem, and that all relevant properties are captured by the rules of
the proof system. Therefore, if the problem is unsatisfiable,
eventually our rules will discover it.

% The outlined proof relies on the theorem on existence of a model.
% In Section~\ref{sec:modelC} we prove the
% theorem on existence of a model, Theorem~\ref{th:model}, for the constant
% domain case; in Section~\ref{sec:modelE} we refine this reasoning for the
% expanding domain case. We conclude the proof of completeness in
% Section~\ref{sec:compProof}.
% \subsection{Constant domain case}\label{sec:modelC}
% %
% In order to prove completeness of the temporal resolution method, we
First, we
introduce additional concepts.
%
%\begin{definition}[colour scheme]
% Let $\{P_i(x)\implies\next
% Q_i(x)\}_{i=1}^{M}$ be the set of \emph{all} non-ground step clauses and
% $\{p_j(x)\implies\next q_j(x)\}_{j=1}^{m}$ be the set of \emph{all} ground step
% clauses. Let $\{\sometime L_k(x)\}$ be the set of \emph{all} non-ground
% eventualities.
Let $\TProb = \langle \uP, \iP, \sP, \eP\rangle$ be a monodic temporal problem.
Let $\{P_1,\dots, P_N\}$ and $\{p_1,\dots,p_n\}$, $N,n\geq0$, be the sets of
all (monadic) predicate symbols and all propositional symbols, respectively,
occurring in $\sP\cup\eP$.

A \emph{predicate colour} $\gamma$ is a set of unary literals such that for
every $P_i(x)\in\{P_1(x),\dots, P_N(x)\}$, either $P_i(x)$ or $\lnot P_i(x)$
belongs to $\gamma$.
A \emph{propositional colour} $\theta$ is a
sequence of propositional literals such that for every
$p_i\in\{p_1,\dots,p_n\}$, either $p_i$ or $\lnot p_i$ belongs to $\theta$.
%
%Let $\Delta$ be the set of all mappings from
%$\{1,\ldots,N\}$ to $\{0,1\}$, and $\Theta$ be the set of all mappings from
%$\{1,\ldots,n\}$ to $\{0,1\}$. An element $\delta \in \Delta$ ($\theta \in
%\Theta$) is represented by the sequence
%$[\delta(1), \ldots, \delta(N)] \in \{0,1\}^N$
%($[\theta(1), \ldots, \theta(n)] \in \{0,1\}^n$).  Let us call elements of
%$\Delta$ and $\Theta$ predicate and propositional \emph{colours}, respectively.
%Let $\Gamma$ be a subset of $\Delta$, $\theta$ be an element of $\Theta$,
Let $\Gamma$ be a predicate colour, $\theta$ be a propositional colour, and
$\rho$ be a map from the set of constants, $\const(\TProb)$, to  $\Gamma$.  A
triple $(\Gamma,\theta, \rho)$ is called a \emph{colour scheme}, and $\rho$ is
called a \emph{constant distribution}.
%%\end{definition}
%%
%If a predicate $P_i(x)$ from $\sP\cup\eP$ ``occurs'' in a  predicate colour
%$\gamma$ (that is, $\gamma(i) = 1$), we also write
%$P_i(x)\in\gamma$; and
%if $\gamma(i) = 0$, we also write $P(x)\notin\gamma$ or $\lnot P(x)\in\gamma$.
%\label{page:occurs} The same convention is used for propositional
%colours and constant distributions.\looseness=-1

%\begin{definition}[categorical formulas]
\begin{note}
The notion of colour scheme came, of course, from the well known
concept used in the decidability proof for the monadic class in
classical first-order logic (see, for example,~\cite{BGG97}). In our
case, $\Gamma$ is the quotient domain (a subset of all possible
equivalence classes of predicate values), $\theta$ is a propositional
valuation, and $\rho$ is a standard interpretation of constants in the
domain $\Gamma$.  We construct quotient structures based only on the
predicates and propositions which occur in the temporal part of the
problem, since only these symbols are really responsible for the
satisfiability (or unsatisfiability) of temporal constraints. In
addition, we have to consider so-called constant distributions
because, unlike in the classical case, we cannot eliminate constants
replacing them by existentially bound variables since in doing this
the monodicity property would be lost.

\end{note}

For every colour scheme $\mathcal{C} = \auf\Gamma,\theta,\rho\zu$ let us
construct the formulae $\mathcal{F}_{\mathcal{C}}$,
$\mathcal{A}_{\mathcal{C}}$, $\mathcal{B}_{\mathcal{C}}$ in the following way.
For every $\gamma \in \Gamma$ and for every $\theta$,
introduce the conjunctions:
$$
\begin{array}{lll}
F_{\gamma}(x) = \bigwedge\limits_{L(x)\in\gamma}L(x);& \quad &
F_{\theta} =
\bigwedge\limits_{l\in\theta}l. \\
\end{array}
$$
Let
$$
\begin{array}{c}
\begin{array}{lcl}
A_{\gamma}(x) &=&
\bigwedge\{L(x) \;|\; {L(x)\implies\next M(x)\in\sP,\  L(x)\in{\gamma}}\},\\
B_{\gamma}(x) &=&
\bigwedge\{M(x) \;|\; {L(x)\implies\next M(x)\in\sP,\  L(x)\in{\gamma}}\},\\
%\end{array}\\
%\begin{array}{lcllcl}
A_{\theta} &=&
\bigwedge\{l \;|\; {l\implies\next m\in\sP,\  l\in{\theta}}\},\\
B_{\theta} &=&
\bigwedge\{m \;|\; {l\implies\next m\in\sP,\  l\in{\theta}}\}.
\end{array}
\end{array}
$$
%
%
%
% $$
% \begin{array}{lll}
% F_{\gamma}(x) =
% \bigwedge\limits_{i \leq N,\; \gamma(i)=1 }P_i(x)\ \andl
%  \bigwedge\limits_{ i \leq N, \; \gamma(i)=0} \notl P_i(x), & \quad &
% F_{\theta} =
% \bigwedge\limits_{i \leq n,\; \theta(i)=1}p_i\ \andl
% \bigwedge\limits_{i \leq n,\; \theta(i)=0} \notl p_i. \\
% \end{array}
% $$
% Let us define two sets of indexes \\
% \hphantom{M}$
% J_{\gamma} = \{i,\; 1\leq i \leq N  \;|\; \gamma(i)=1
% \text{ and $P_i(x)\implies\next M_i(x)$ belongs to $\sP$ for some $M_i$}\}$ and\\
% \hphantom{M}$ J_{\theta} = \{j, \; 1\leq i \leq n\;|\; \theta(j)=1
% \text{ and $p_j\implies\next m_i$ belongs to $\sP$ for some $m_i$}\}.
% $\\
(Recall that there are no two different step clauses with the same left-hand
side.)
% Let $
% \begin{array}[t]{lll}
% A_{\gamma}(x) =
% \bigwedge\limits_{i\in J_{\gamma}}P_i(x),  & \quad &
% B_{\gamma}(x) =
% \bigwedge\limits_{i\in J_{\gamma}}M_i(x),
% \end{array}
% \quad
% \begin{array}[t]{lll}
% A_{\theta} =
% \bigwedge\limits_{i\in J_{\theta}}p_i, & \quad &
% B_{\theta} =
% \bigwedge\limits_{i\in J_{\theta}}m_i.
% \end{array}
% $

\noindent Now $\mathcal{F}_{\mathcal{C}}$, $\mathcal{A}_{\mathcal{C}}$,
$\mathcal{B}_{\mathcal{C}}$ are of the following forms:
$$
\begin{array}{l}
\mathcal{F}_{\mathcal{C}} = \bigwedge\limits_{\gamma \in \Gamma}
\exists x F_{\gamma}(x) \andl
F_{\theta} \andl
\bigwedge\limits_{c \in C} F_{\rho(c)}(c)  \andl
\forall x \bigvee\limits_{\gamma \in \Gamma} F_{\gamma}(x),\\
\begin{array}{l}
\mathcal{A}_{\mathcal{C}} = \bigwedge\limits_{\gamma \in \Gamma}
\exists x A_{\gamma}(x)  \andl A_{\theta} \andl
\bigwedge\limits_{c \in C} A_{\rho(c)}(c)  \andl
\forall x \bigvee\limits_{\gamma \in \Gamma} A_{\gamma}(x),
\end{array}\\
\mathcal{B}_{\mathcal{C}} = \bigwedge\limits_{\gamma \in \Gamma}
\exists x B_{\gamma}(x)  \andl B_{\theta} \andl
\bigwedge\limits_{c \in C} B_{\rho(c)}(c)  \andl
\forall x \bigvee\limits_{\gamma \in \Gamma} B_{\gamma}(x).
\end{array}
$$
We can consider the formula $\mF_{\mC}$ as a ``categorical'' formula
specification of the quotient structure given by a colour scheme. In
turn, the formula $\mA{_\mC}$ represents the part of this
specification which is ``responsible'' just for ``transferring''
requirements from the current world (quotient structure) to its
immediate successors, and $\mB_{\mC}$ represents the result of transferal.
%\end{definition}
%
\begin{example}\label{ex:graph-1}
Consider a monodic temporal problem, $\TProb$, given by\\
$$ 
\begin{array}{lcllcl}
  \iP & = & \emptyset, \qquad&
  \sP & = & \{P(x)\implies\next P(x)\}, \\
  \uP & = & \{ l\implies \exists x P(x)\}, \qquad &
  \eP & = & \{\sometime\lnot P(x), \sometime l \}.
\end{array}
$$
For this problem, there exist two predicate colours, 
$\gamma_1 = [P(x)]$ and
$\gamma_2 = [\lnot P(x)]$; 
two propositional colours
$\theta_1 = [l]$ and
$\theta_2 = [\lnot l]$;
and six colour schemes, 
$$
\begin{array}{lcllcl}
\mC_1 & = & (\{\gamma_1\}, \theta_1), \qquad&
\mC_4 & = & (\{\gamma_1\}, \theta_2), \\
\mC_2 & = & (\{\gamma_2\}, \theta_1), \qquad &
\mC_5 & = & (\{\gamma_2\}, \theta_2), \\
\mC_3 & = & (\{\gamma_1, \gamma_2\}, \theta_1),\qquad &
\mC_6 & = & (\{\gamma_1, \gamma_2\}, \theta_2).
\end{array}
$$
The categorical formulae for these colour schemes are the following:
%\noindent Now, we construct the formulae:
$$
\begin{array}{lll}
\mF_{\mC_1} = \exists x P(x)\land\forall x P(x)\land l\qquad& 
\mA_{\mC_1} = \exists x P(x)\land\forall x P(x)\;& 
\mB_{\mC_1} = \exists x P(x)\land\forall x P(x)\\
\mF_{\mC_2} = \exists x \lnot P(x)\land\forall x \lnot P(x) \land l&
\mA_{\mC_2} = \true& 
\mB_{\mC_2} = \true\\
\mF_{\mC_3} = \exists x P(x)\land\exists x \lnot P(x) \land l&
\mA_{\mC_3} = \exists x P(x)&
\mB_{\mC_3} =\exists x P(x) \\
\mF_{\mC_4} = \exists x P(x)\land\forall x P(x) \land \lnot l& 
\mA_{\mC_4} = \exists x P(x)\land\forall x P(x)\ & 
\mB_{\mC_4} = \exists x P(x)\land\forall x P(x)\\
\mF_{\mC_5} = \exists x \lnot P(x)\land\forall x \lnot P(x) \land \lnot l\ &
\mA_{\mC_5} = \true& 
\mB_{\mC_5} = \true\\
\mF_{\mC_6} = \exists x P(x)\land\exists x \lnot P(x) \land \lnot l&
\mA_{\mC_6} = \exists x P(x)&
\mB_{\mC_6} =\exists x P(x) 
\end{array}
$$
\end{example}

\begin{definition}[Canonical \moreboris{Merged \Grounded{}}
Step Clauses]
\noindent Let $\TProb$ be a first-order temporal problem, $\mC$ be a
colour scheme for $\TProb$. Then the clause
$$
  (\mA_{\mC} \imp \next \mB_{\mC}),
$$
%% where $\mA_{\mC}$ and $ \mB_{\mC}$ are defined as above
is called a
%\emph{canonical divided merged step clause} for $\TProb$.
\emph{canonical merged \grounded{} step clause} for $\TProb$.

If all the sets $J_{\gamma}$, for all $\gamma\in\Gamma$, and $J_{\theta}$
are empty, the clause $(\mA_{\mC} \imp \next \mB_{\mC})$ \emph{degenerates}
to $(\ltrue
\imp \next \ltrue)$. If a conjunction $A_{\gamma}(x)$, $\gamma \in \Gamma$,
is empty, that is its truth value is $\ltrue$, then the formula
$\forall x \bigvee_{\gamma \in \Gamma} A_{\gamma}(x)$
(or $\forall x \bigvee_{\gamma \in \Gamma} B_{\gamma}(x)$)
disappears from $\mathcal{A}_{\mathcal{C}}$ (or from
$\mathcal{B}_{\mathcal{C}}$ respectively). In the propositional case, the  clause
$(\mA_{\mC} \imp \next \mB_{\mC})$
reduces to $(A_{\theta} \imp \next B_{\theta})$.
\end{definition}
\begin{definition}[Canonical Merged Step Clause]
Let $\mC$ be a colour scheme,
$
\mA_{\mC}\implies\next\mB_{\mC}
$
%be a canonical divided merged step clause, and $\gamma\in\mC$.
be a canonical  merged \grounded{} step clause, and $\gamma\in\mC$.
$$
\forall x(\mA_{\mC}\land A_{\gamma}(x)\implies\next(\mB_{\mC}\land B_{\gamma}(x)))
$$
is called a \emph{canonical merged step clause}. If the set $J_{\gamma}$ is
empty, the truth value of the conjunctions $A_{\gamma}(x)$, $B_{\gamma}(x)$
is $\true$, and the canonical merged step clause is just a
%canonical divided merged step clause.
canonical merged \grounded{} step clause. {$\gamma\in\mC$ abbreviates here $\gamma\in\Gamma$, where $\mC = (\Gamma,\theta,\rho)$.}
\end{definition}
\begin{definition}[Behaviour Graph]
Now, given a temporal problem $\TProb = \langle \uP, \iP,
\sP,\eP\rangle$ we define a finite directed graph $G$ as
follows. Every vertex of $G$ is a colour scheme $\mC$ for $\TProb$ such
that  $\uP \cup \mathcal{F}_{\mathcal{C}}$ is satisfiable. For
each vertex $\mC = (\Gamma,\theta,\rho)$, there is an edge in $G$ to
 $\mC' = (\Gamma',\theta',\rho')$, if $ ~\uP \andl
\mathcal{F}_{\mathcal{C}'} \andl \mB_{\mathcal{C}}$ is satisfiable.
They are the only edges originating from $\mC$.

\noindent A vertex $\mathcal{C}$ is designated as an \emph{initial} vertex
of $G$ if $\iP \andl \uP \andl \mathcal{F}_{\mathcal{C}} $ is
satisfiable.

\noindent The \emph{behaviour graph} $H$ of $\TProb$ is the subgraph of
$G$ induced by the set of all vertices reachable from the initial vertices.
\end{definition}
%%%%%%%%%%%%%%%%%%%%%%%%%%%%%%%%%%%%%%%%%%%%%%%%%%%%%%%%%%%%%%%%%%%%%%%%%%%%%%
\begin{figure}[tb]
\begin{center}
\psfrag{I0}{\scriptsize$\!\mC_1$}
\psfrag{I1}{\scriptsize$\!\mC_3$}
\psfrag{I2}{\scriptsize$\!\mC_4$}
\psfrag{I3}{\scriptsize$\!\mC_6$}
\psfrag{I4}{\scriptsize$\!\mC_5$}
\includegraphics[width=0.50\textwidth]{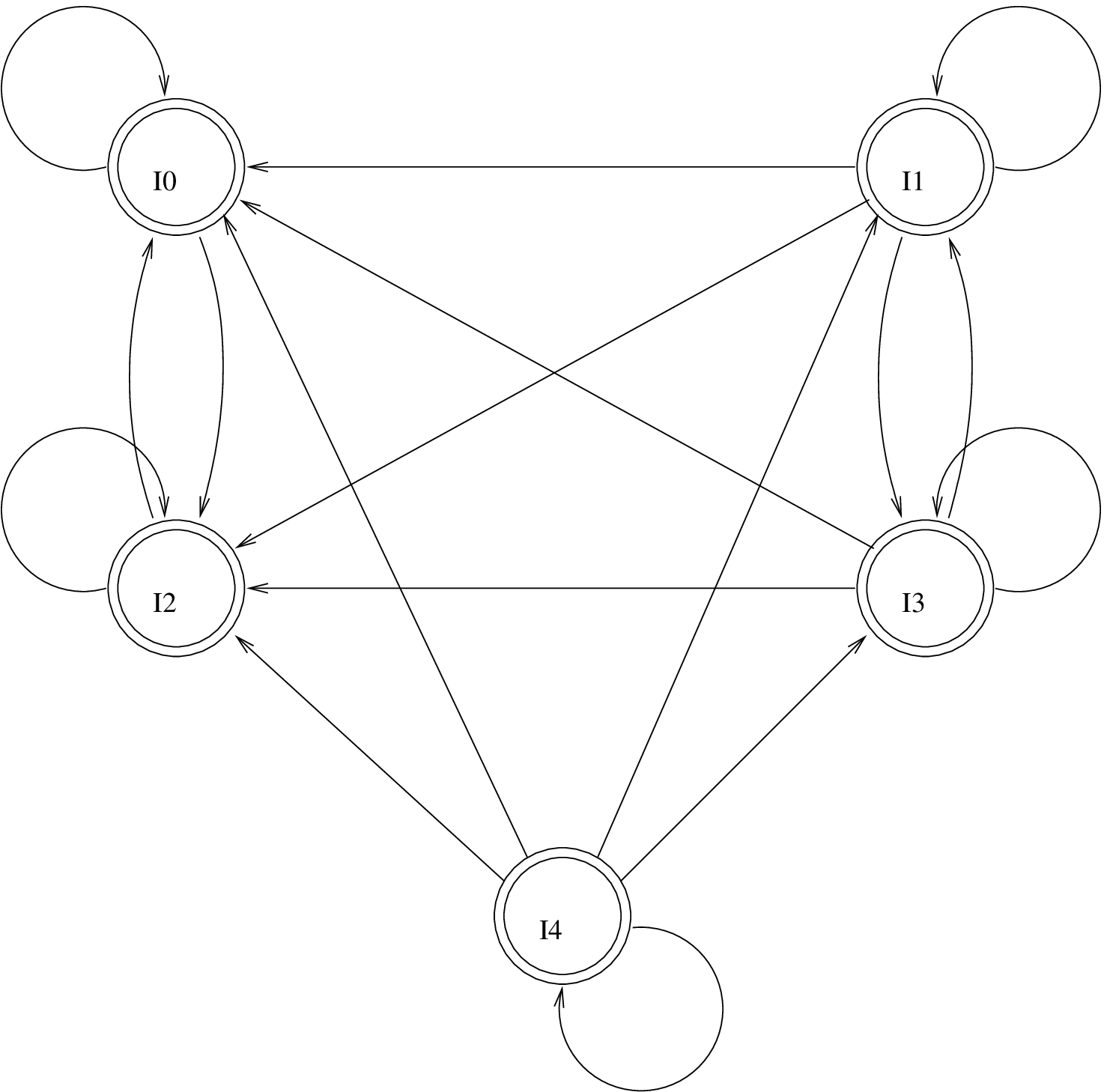}
\caption{
Behaviour graph for the problem
$ \iP = \emptyset $,
$ \uP = \{ l\implies \exists x P(x)\} $,
$ \sP = \{ P(x)\implies \protect\next P(x) \} $,
$ \eP = \{\sometime\lnot P(x), \sometime l \}$
(Example~\ref{example:graph}).
}\label{fig:graph}
\end{center}
\end{figure}
%%%%%%%%%%%%%%%%%%%%%%%%%%%%%%%%%%%%%%%%%%%%%%%%%%%%%%%%%%%%%%%%%%%%%%%%%%%%%%
\begin{example}[Example~\ref{ex:graph-1} contd.]\label{example:graph}
Let us construct the behaviour graph for the problem given in
Example~\ref{ex:graph-1}.  Note that $\mF_{\mC_2} \land \uP \models \perp$, 
so the vertex $\mC_2$ is not in the graph.
The behaviour graph for $\TProb$, given in Fig.~\ref{fig:graph}, consists of
five vertices; all of them are initial.

There is an edge in the graph from the node $\mC_3$ to the node $\mC_1$ since
the formula  $\uP\land \mF_{\mC_1}\land \mB_{\mC_3}$,
$$
\begin{array}{l}
\underbrace{l\implies\exists x P(x)}_{\uP}
\land \underbrace{\exists x P(x)\land\forall x P(x)\land l}_{\mF_{\mC_1}}
\land \underbrace{\exists x P(x)}_{\mB_{\mC_3}},
\end{array}
$$
is satisfiable. There is no edge from $\mC_1$ to $\mC_3$ since the 
formula $\uP\land \mF_{\mC_3}\land \mB_{\mC_1}$,
$$
\begin{array}{l}
\underbrace{l\implies\exists x P(x)}_{\uP}
\land \underbrace{\exists x P(x)\land\exists x \lnot P(x) \land l}_{\mF_{\mC_3}}
\land \underbrace{\exists x P(x)\land\forall x P(x)}_{\mB_{\mC_1}}
\end{array}
$$
is unsatisfiable. Other edges are considered in a similar way.
\end{example}

\begin{lemma}
\label{lem: incl}%
Let $\TProb_1 = \langle \uP_1,\iP,\sP,\eP\rangle$ and  $\TProb_2 =
\langle \uP_2,\iP,\sP, \eP\rangle$ be two \moreboris{problems}
such that
$\uP_1 \subseteq \uP_2$. Then the behaviour graph of \moreboris{$\TProb_2$}
is a subgraph of the behaviour graph  of \moreboris{$\TProb_1$}.%
\end{lemma}
\begin{proof}
Satisfiability of $\uP_2$ implies satisfiability of $\uP_1$.
\end{proof}
\begin{definition}[Path; Path Segment]
A \emph{path}, $\pi$, through a behaviour graph, $H$, is a function from
$\Nat$ to the vertices of the graph such that for any $i\geq0$ there is an edge
$\langle \pi(i),\pi(i+1) \rangle$ in $H$. In a similar way, we define a
\emph{path segment} as a function from $[m,n]$, $m<n$, to the vertices of $H$
with the same property.
%the pair of
%vertices $\langle\pi(i),\pi(i+1)\rangle$ is adjacent
\end{definition}
Recall that vertices of the behaviour graph of a problem, $\TProb$,
are quotient representations of ``intermediate'' interpretations $\gM_n$
in possible models of $\TProb$. Intuitively, if a pair of vertices, or
of colour schemes, $\mC$ and $\mC'$ is suitable, then this pair can
represent adjacent interpretations $\gM_i$ and $\gM_{i+1}$ in a model
of $\TProb$.  The definition of predicate colour suitability given
below expresses the condition when a pair of predicate colours specify
an element in adjacent interpretations with regard to the step part of
$\TProb$. A similar intuition is behind the notions of suitable
propositional colours and suitable constant distributions.
%
%Intuitive meaning of an edge  from a vertex $\mC$ to a vertex
%$\mC$

\begin{definition}[Suitability]
For $\mC = ( \Gamma, \theta, \rho )$ and $\mC' = ( \Gamma', \theta',
\rho' )$, let $(\mC, \mC')$ be an ordered pair of colour schemes for \moreboris{a temporal problem}
$\TProb$.

\noindent An ordered pair of predicate colours $(\gamma, \gamma\,')$
where $\gamma \in \Gamma$, $\gamma\,' \in \Gamma'$ is called
\emph{suitable} if the formula $\mU \andl \exists x (F_{\gamma\,'}(x)
\andl B_{\gamma}(x))$ is satisfiable;

\noindent Similarly, an ordered pair of
propositional colours $(\theta, \theta')$ is suitable if $\mU \andl
F_{\theta'} \andl B_{\theta}$ is satisfiable; and

\noindent an ordered pair of
constant distributions $(\rho,\rho')$ is suitable if, for every $c \in
C$, the pair $(\rho(c),\rho'(c))$ is suitable.
\end{definition}
Note that the satisfiability of $\exists x(F_{\gamma\,'}(x) \andl
B{_\gamma}(x))$ implies $\models\forall x(F_{\gamma\,'}(x)\implies
B{_\gamma}(x))$ as the conjunction $F_{\gamma\,'}(x)$ contains a
valuation at $x$ of \emph{all} predicates occurring in
%% the expression
$B{_\gamma}(x)$.
\begin{lemma}\label{lem:lemma3}
Let $H$ be the behaviour graph for the problem
$\TProb =\, \langle\uP, \iP, \sP, \eP\rangle$ with an edge from a vertex
$\mC=(\Gamma, \theta,\rho)$ to a vertex $\mC'=(\Gamma', \theta',\rho')$.
Then
\begin{enumerate}
\item for every $\gamma\in\Gamma$ there exists a $\gamma\,'\in\Gamma'$ such that
the pair $(\gamma,\gamma\,')$ is suitable;
\item for every $\gamma\,'\in\Gamma'$ there exists a $\gamma\in\Gamma$ such that
the pair $(\gamma,\gamma\,')$ is suitable;
\item the pair of propositional colours $(\theta,\theta')$ is suitable;
\item the pair of constant distributions $(\rho,\rho')$ is suitable.
\end{enumerate}
\end{lemma}
\begin{proof} From the definition of a behaviour graph it follows that
$ \uP \andl \mathcal{F}_{\mathcal{C}'} \andl \mB_{\mC}$ is satisfiable.
Now to prove the first item it is enough to note that satisfiability of the
expression  $ \uP \andl \mathcal{F}_{\mathcal{C}'} \andl \mB_{\mC}$ implies
satisfiability of $\mU \andl (\forall x \bigvee\limits_{\gamma' \in \Gamma'}
F_{\gamma'}(x))
\andl \exists x B_{\gamma}(x)$. This, in turn, implies satisfiability of its
logical
consequence $\mU \andl \bigvee\limits_{\gamma' \in \Gamma'} \exists x
(F_{\gamma'}(x) \andl B_{\gamma}(x) )$. So, one of the members of this
disjunction must be satisfiable.  The second item follows from the
satisfiability of $\mU \andl (\forall x \bigvee\limits_{\gamma \in
\Gamma} B_{\gamma}(x))
\andl \exists x F_{\gamma'}(x)$. Other items are similar.
\end{proof}
\begin{example}[Example~\ref{example:graph} cont.]
\label{ex:graph-3}
Let us consider suitability of predicate and propositional colours from
Example~\ref{ex:graph-1}.

\noindent Since the formula $ \mU \andl \exists x (F_{\gamma_1}(x) \andl
B_{\gamma_2}(x))$, where $\uP = \{l\implies\exists x P(x)\}$, $F_{\gamma_1} =
P(x)$, and $B_{\gamma_2} = \true$, is satisfiable, the pair 
$(\gamma_1, \gamma_2)$ is suitable.

\noindent Since the formula $ \mU \andl \exists x (F_{\gamma_2}(x) \andl
B_{\gamma_1}(x))$, where $\uP = \{l\implies\exists x P(x)\}$, $F_{\gamma_2} =
\lnot P(x)$, and $B_{\gamma_1} = P(x)$, is unsatisfiable, the pair 
$(\gamma_2, \gamma_1)$ is not suitable.

\noindent In a similar way, it can be easily checked that the
pairs of predicate colours
$$
(\gamma_1, \gamma_1) \quad\textrm{and}\quad
(\gamma_2, \gamma_2), \quad
$$
and the pairs of propositional colours
$$
(\theta_1,\theta_1),\quad
(\theta_1,\theta_2),\quad
(\theta_2,\theta_1),\textrm{ and}\quad
(\theta_2,\theta_2)
$$
are suitable.
\end{example}

Let $H$ be the behaviour graph for a temporal problem
$\TProb = \langle \uP,\iP,\sP, \eP\rangle$ and
$
\pi = \mC_0, \ldots, \mC_n, \ldots $ be a path in $H$ where $\mC_i
= (\Gamma_i,\theta_i,\rho_i)$.  Let $ \mG_0 = \iP \cup
\{\mF_{\mC_o}\}$ and $ \mG_n = \mF_{\mC_n} \andl \mB_{\mC_{n-1}}$ for
$n \geq 1$. According to the definition of a behaviour graph,
the set $\mU \cup \{\mG_n\}$ is satisfiable for every $n \geq 0$.

From classical model theory, since the language $\mL$ is
countable and does not contain equality, the following lemma holds.
\begin{lemma}
\label{lem:lemma4}
Let $\kappa$ be a cardinal, $\kappa\geq\aleph_0$. For every $n\geq 0$,
if the set $\uP\cup\{\mG_n\}$ is satisfiable then there exists an
$\mL$-model $\gM_n=\langle D,I_n \rangle$ of $\uP\cup\{\mG_n\}$ such
that for every $\gamma\in\Gamma_n$ the set $D_{(n,\gamma)} = \{a\in D
\;|\; \gM_n\models F_{\gamma}(a)\}$ is of cardinality $\kappa$.
\end{lemma}
\begin{definition}[Run/E-Run]
Let $\pi$ be a path through a behaviour graph $H$ \moreboris{of a temporal problem $\TProb$, and $\pi(i) = (\Gamma_i, \theta_i,\rho_i)$%
%=\TSpec\union\eP$
}.  By a \emph{run} in
$\pi$ we mean a function $r(n)$ from $\Nat$ to $\bigcup_{i \in \Nat}
\Gamma_i$ such that for every $n \in \Nat$, $r(n) \in \Gamma_n$ and
the pair $(r(n),r(n+1))$ is suitable.  In a similar way, we define a
\emph{run segment} as a function from
$[m,n]$, $m<n$, to $\bigcup_{i \in \Nat} \Gamma_i$ with the same
property.

A run $r$ is called an \emph{e-run} if for all $i\geq 0$ and
for every non-ground eventuality $\sometime L(x)\in\eP$ there
exists $j>i$ such that $L(x)\in r(j)$.
\end{definition}
Let $\pi$ be a path, the set of all runs in $\pi$ is denoted by
$\mR(\pi)$, and the set of all e-runs in $\pi$ is denoted by
$\mR_{e}(\pi)$. If $\pi$ is clear, we may omit it.
\begin{example} $\pi = \mC_3,\mC_6,\mC_3,\mC_6,\dots$ is a path through the
behaviour graph given in Fig.~\ref{fig:graph}.
$r_1 = \gamma_1, \gamma_1, \gamma_1,\dots$ and $r_2 =
\gamma_1,\gamma_2,\gamma_1,\gamma_2,\dots$ are both runs in $\pi$.
$r_2$ is an e-run, but $r_1$ is not.
\end{example}
%\end{definition}
%It follows from the definition of $H$ that for every $c \in C$ the
%function $r_c$ defined by $r_c(n) = \rho_n(c)$ is a run in $\pi$.
%
We now relate properties of the behaviour graph for a problem to the
satisfiability of the problem.
\begin{theorem}[Existence of a model]\label{th:model}
Let $\TProb=\langle \uP,\iP,\sP,\eP\rangle$ be a temporal problem. Let
$H$ be the behaviour graph of $\TProb$, let $\mC$ and $\mC'$ be
vertices of $H$ such that $\mC = (\Gamma,\theta,\rho)$ and $\mC' =
(\Gamma',\theta',\rho')$. If both the set of initial vertices of $H$
is non-empty and  the following conditions hold
\begin{enumerate}
\item\label{eq:model1}
For every vertex $\mC$, predicate colour $\gamma\in\Gamma$,  and
non-ground eventuality $\sometime L(x)\in\eP$ there exist
a vertex $\mC'$ and a predicate colour $\gamma'\in\Gamma'$
such that
$$
\left(\left(\mC,\gamma\right)\to^{+}\left(\mC',\gamma\,'\right)
      \land L\left(x\right)\in\gamma\,'\right);
$$
\item\label{eq:model2} For every vertex $\mC$, constant $c\in\const(\TProb)$, and
non-ground eventuality $\sometime L(x)\in\eP$,
there exists a vertex $\mC'$  such that
$$
\left(\mC\to^{+}\mC'\land L\left(x\right)\in\rho'\left(c\right)\right);
$$
\item\label{eq:model3} For every vertex $\mC$ and ground eventuality
$\sometime l\in\eP$, there exists a vertex $\mC'$ such that
$$
\left(\mC\to^{+}\mC' \land l\in\theta'\right)
$$
\end{enumerate}
 then $\TProb$ has a model.
Here $(\mC,\gamma)\to^{+}(\mC',\gamma\,')$ denotes that there
exists a path $\pi$ from $\mC$ to $\mC'$ such that $\gamma$ and
$\gamma\,'$ belong to a run in $\pi$; and $\mC\to^{+}\mC'$ denotes that
there exists a path from $\mC$ to $\mC'$.
\end{theorem}
The proof proceeds as follows. First, we provide a lemma showing that,
under the conditions of Theorem~\ref{th:model}, there exists a path
through the behaviour graph satisfying certain properties, and then we
show that, if such a path exists, then the problem has a model.
\begin{lemma}\label{lem:path}
Under the conditions of Theorem~\ref{th:model}, there exists a path
$\pi$ through $H$ where:
\begin{enumerate}
\item[(a)] $\pi(0)$ is an initial vertex of $H$;
\item[(b)] for every colour scheme $\mC = \pi(i),\; i\geq0$, and
every ground eventuality literal $\sometime l\in\eP$ there exists a colour
scheme $\mC' = \pi(j)$, $j > i$,  such that $l\in\theta'$;
\item[(c)] for every colour scheme $\mC = \pi(i),\; i\geq0$ and
every predicate colour $\gamma$ from the colour scheme there exists
an e-run $r\in\mR_{e}(\pi)$ such that $r(i) = \gamma$; and
%
%\item For every colour scheme $\mC = \pi(i),\; i\geq0$, every predicate colour
%$\gamma$ from the colour scheme, every eventuality literal $L(x)\in\eP$,
%and for every run $r'\in\mR_{e}$  such that $r'(i) = \gamma$ there exists
%$j\geq i$ such that $r'(j) = \gamma\,'$ and $L(x)\in\gamma\,'$;
%
\item[(d)] for every constant $c\in \mL$, the function $r_c(n)$
defined by $r_c(n) = \rho_n(c)$, where $\rho_n$ is the constant
distribution from $\pi(n)$, is an  e-run in $\pi$.
\end{enumerate}
\end{lemma}
\begin{proof}[of Lemma~\ref{lem:path}]
Let $\sometime L_1(x),\dots,\sometime L_k(x)$ be all non-ground eventuality
literals from $\eP$; $\sometime l_1,\dots,\sometime l_p$ be all ground
eventuality literals from $\eP$; and $c_1,\dots,c_q$ be all constants
of $\TProb$. Let $\mC_0$ be an initial vertex of $H$. We construct the
path $\pi$ as follows. Let $\{\gamma_{1},\dots,\gamma_{s_0}\}$ be all
predicate colours from $\Gamma_{\mC_0}$. By condition
(\ref{eq:model1}) there exists a vertex $\mC_0^{(\gamma_1,L_1)}$ and a
predicate colour $\gamma_{1}^{(1)}\in\Gamma_{\mC_0^{(\gamma_1,L_1)}}$
such that $(\mC_0,\gamma_{1})\to^+(\mC_0^{(\gamma_1,L_1)},
\gamma_{1}^{(1)})$ and $L_1(x)\in\gamma_{1}^{(1)}$. In the same way,
there exists a vertex $\mC_0^{(\gamma_1,L_2)}$ and a predicate colour
$\gamma_{1}^{(2)}\in\Gamma_{\mC_0^{(\gamma_1,L_2)}}$ such that
$(\mC_{0}^{(\gamma_1,L_1)},\gamma_{1}^{(1)})\to^+(\mC_0^{(\gamma_1,L_2)},
\gamma_{1}^{(2)})$ and $L_2(x)\in\gamma_{1}^{(2)}$. And so
on. Finally, there exists a vertex $\mC_0^{(\gamma_1,L_k)}$ and a
predicate colour $\gamma_{1}^{(k)}\in\Gamma_{\mC_0^{(\gamma_1,L_k)}}$
such that
$(\mC_{0}^{(\gamma_1,L_{k-1})},\gamma_{1}^{(k-1)})\to^+(\mC_0^{(\gamma_1,L_k)},
\gamma_{1}^{(k)})$ and $L_k(x)\in\gamma_{1}^{(k)}$. Clearly,
$\gamma_{1}$,
\dots,$\gamma_{1}^{(1)}$,\dots,$\gamma_{1}^{(2)}$,\dots,$\gamma_{1}^{(k)}$
forms a segment of a run and every non-ground eventuality is satisfied
along this segment.

Now, let $\gamma_{2}^{(0)}$ be any successor of $\gamma_{2}$ in
$\Gamma_{\mC_{0}^{(\gamma_1,L_k)}}$. As above, there exists a sequence
of vertices $\mC_0^{(\gamma_2,L_1)}$,\dots, $\mC_{0}^{(\gamma_2,L_{k})}$
and a sequence of predicate colours
$\gamma_{2}^{(1)}\in\Gamma_{\mC_{0}^{(\gamma_2,L_1)}}$,\dots,
$\gamma_{2}^{(k)}\in\Gamma_{\mC_{0}^{(\gamma_2,L_k)}}$ such that
$\gamma_{2}$,\dots,$\gamma_{2}^{(0)},\dots,\gamma_{2}^{(1)},\dots,\gamma_{2}^{(k)}$
forms a segment of a run and every non-ground eventuality is satisfied along
this segment. Continue this construction. At a certain point we construct a segment
of a path from $\mC_0$ to a vertex $\mC_0^{(\gamma_{s_0},L_k)}$ such
that
%condition (\ref{eq:model1})
%is satisfied on this segment for $\mC = \mC_0$. Moreover,
for every $\gamma\in\mC_0$ there exists
$\gamma\,'\in\mC_0^{(\gamma_{s_0}, L_k)}$ such that all
eventualities are satisfied on the run-segment from $\gamma$ to
$\gamma\,'$.

In a similar way we can construct a vertex $\mC_0^{(c_1,L_1)}$ such
that $\mC_0^{(\gamma_{s_0},L_k)}\to^+\mC_0^{(c_1,L_1)}$ and
$L_1(x)\in\rho_{\mC_0^{(c_1,L_1)}}(c_1)$. And so on.
Then we can construct a vertex $\mC_0^{(l_1)}$ such that
$\mC_0^{(c_q,L_k)}\to^+\mC_0^{(l_1)}$ and
$l_1\in\theta_{\mC_0^{(l_1)}}$. And so on.

Finally, we construct a vertex $\mC_0' = \mC_0^{(l_p)}$ such that
$\mC_0\to^+\mC_0'$ and on this path segment all conditions of the
theorem hold for $\mC=\mC_0$. Let us denote this path segment as
$\lambda_0$, and let $\mC_1$ be any successor of $\mC_0'$.

By analogy, we can construct a vertex $\mC_1'$ and a path segment
$\lambda_1$ from $\mC_1$ to $\mC_1'$ such that all conditions of the
theorem hold for $\mC=\mC_1$. An so forth. Eventually, we construct a
sequence $\mC_0$, $\mC_1$,\dots, $\mC_j$ such that there exists
$n,\,0\leq n<j$ and $\mC_n = \mC_j$ because there are only finitely
many different colour schemes. Let $\pi_1 =
\lambda_0,\dots,\lambda_{n-1}$, $\pi_2 =
\lambda_n,\dots\lambda_{j-1}$. Now, we define our path $\pi$ as
$\pi_1(\pi_2)^*$. Properties (a) and (b) evidently hold on $\pi$.

Let $\mC = \pi(i)$ and $\gamma\in\Gamma_{\mC}$. Clearly, there exist
$\gamma\,'\in\mC_{0}$ and $\gamma\,''\in\mC_{n}$ such that
$(\mC_0,\gamma\,')\to^+(\mC,\gamma)$ and
$(\mC,\gamma)\to^+(\mC_{n},\gamma\,'')$. Since for every
$\gamma\,''\in\mC_n$ there exists $\gamma\,'''\in\mC_n^{(\gamma_{s_n},
L_k)}$ such that all eventualities are satisfied on the run-segment
from $\gamma\,''$ to $\gamma\,'''$ and there exists
$\gamma^{(4)}\in\mC_n$, $(\mC_n^{(\gamma_{s_n}, L_k)},
\gamma\,''')\to^+(\mC_n,\gamma^{(4)})$, then there is an e-run, $r$,
such that $r(i) = \gamma$, that is, property (c) holds.

Note that, for every constant $c$ of $\TProb$ the sequence $r_c(n)$ is
a run in $\pi$. By construction, for every $\sometime L(x)\in\eP$
there is a vertex $\mC_n^{(c, L)}$ in $\pi_2$ such that
$L(x)\in\rho_{\mC_n^{(c, L)}}(c)$. Therefore, $r_c(n)$ is an e-run in
$\pi$ and property (d) holds.
\end{proof}
\begin{proof}[of Theorem~\ref{th:model}]
Following~\cite{HWZ00,DF01} take a cardinal $\kappa\geq\aleph_0$ exceeding the
cardinality of the set $\mR_{e}$. Let us define a domain
$D = \{\langle r, \xi \rangle \;|\; r\in\mR_{e}, \xi<\kappa\}$. Then for
every $n\in\Nat$ we have\\{}
\hbox{\qquad} $D = \bigcup\limits_{\gamma\in\Gamma_n}D_{(n,\gamma)}$,
where $D_{(n,\gamma)} = \{\langle r,\xi  \rangle \;|\; r(n) = \gamma\}$ and
$\left| D_{(n,\gamma)} \right| = \kappa$.\\{}
Hence, by Lemma~\ref{lem:lemma4}, for every $n\in\Nat$ there exists an
$\mL$-structure $\gM_n = \langle D, I_n\rangle$ satisfying
$\uP\cup\{\mG_n\}$ such that $D_{(n,\gamma)} = \{\langle r, \xi
\rangle \in D\;|\; \gM_n\models F_{\gamma}(\langle r,\xi\rangle)
\}$. Moreover, we can suppose that $c^{I_n} = \langle r_c, 0 \rangle$
for every constant $c\in \const(\TProb)$.  A potential first order temporal
model is $\gM = \langle D, I \rangle$, where $I(n) = I_n$ for all
$n\in\Nat$. To be convinced of this we have to check validity of step
and eventuality clauses.  (Recall that satisfiability of $\iP$ in
$\gM_0$ is implied by satisfiability of $\mG_0$ in $\gM_0$.)

Let $\always\forall x (P_i(x)\implies\next R_i(x))$ be an arbitrary
step clause; we show that it is true in $\gM$. Namely, we show that
for every $n\geq0$ and every $\langle r, \xi \rangle \in D$, if
$\gM_n\models P_i(\langle r,\xi \rangle)$ then $\gM_{n+1}\models
R_i(\langle r,\xi \rangle)$. Suppose $r(n) = \gamma \in \Gamma_n$ and
$r(n+1) = \gamma\,'\in\Gamma'$, where $(\gamma,\gamma\,')$ is a suitable
pair in accordance with the definition of a run. It follows that
$\langle r,\xi \rangle\in D_{(n,\gamma)}$ and $\langle r,\xi \rangle
\in D_{(n+1, \gamma\,')}$, in other words $\gM_n\models
F_{\gamma}(\langle r,\xi \rangle)$ and $\gM_{n+1}\models
F_{\gamma\,'}(\langle r,\xi \rangle)$. Since $\gM_n\models P_i(\langle
r, \xi \rangle)$ then $\gamma(i) = 1$. It follows that $R_i(x)$ is a
conjunctive member of $B_{\gamma}(x)$. Since the pair
$(\gamma,\gamma\,')$ is suitable, it follows that the conjunction
$\exists x(F_{\gamma\,'}(x)\land B_{\gamma}(x))$ is satisfiable and,
moreover, $\models\forall x (F_{\gamma\,'}(x)\implies
B_{\gamma}(x))$. Together with $\gM_{n+1}\models F_{\gamma\,'}(\langle
r,\xi \rangle)$ this implies that $\gM_{n+1}\models R_i(\langle r,\xi
\rangle)$. Propositional step clauses are treated in a similar way.

% Let $\always \forall x \sometime L(x)$ be an arbitrary eventuality clause.
% We show that for every $n\geq 0$ and every $\langle r, \xi \rangle \in D$,
% there exists $m\geq n$ such that $\gM_m\models L(\langle r, \xi \rangle)$.
% Suppose $r(n) = \gamma \in \Gamma_n$. Then, there exists $m\geq n$ such that
% $r(m) = \gamma\,' \in \Gamma'$ and $L(x) \in \gamma\,'$. It follows that
% $\langle r,\xi \rangle \in D_{(m,\gamma\,')}$, that is
% $\gM_m\models F_{\gamma\,'}(\langle r, \xi \rangle)$. In particular,
% $\gM_m\models L(\langle r, \xi \rangle)$. Propositional eventuality clauses
% are considered in a similar way.
Let $(\always \forall x) \sometime L(x)$ be an arbitrary
eventuality %\marginnote{I changed a bit this paragraph}
clause. We show that for every $n \geq 0$ and every $\langle r, \xi \rangle
\in D$, $r\in\mR_{e}, \xi<\kappa$, there exists $m > n$
% \marginnote{Besides should be $m > n$ in accordance with the
% definition of e-run}
such that $\gM_m\models L(\langle r, \xi
\rangle)$. Since $r$ is an e-run,  there exists
$\mC'=\pi(m)$ for some $m>n$ such that
$r(m) = \gamma\,' \in \Gamma'$ and $L(x) \in \gamma\,'$.
It follows that $\langle r,\xi \rangle \in D_{(m,\gamma\,')}$, that
is $\gM_m\models F_{\gamma\,'}(\langle r, \xi \rangle)$. In
particular, $\gM_m\models L(\langle r, \xi \rangle)$.
Propositional eventuality clauses are considered in a similar way.
\end{proof}

\begin{note}\label{note:3to2}
%Let \TProb{} be a temporal problem. Then, for $\TProb^c$ condition
For \emph{constant flooded} temporal problems condition \ref{eq:model3} of
Theorem~\ref{th:model} implies condition \ref{eq:model2}.
\end{note}
\begin{lemma}\label{lem:csch}
Let $\gM$ be a first-order temporal structure.
Then there exists a colour scheme $\mC$ such that 
$\gM\models\mF_{\mC}$.
\end{lemma}
\begin{proof}
Let $\gM = \langle D, I\rangle$.
For every $a\in D$,
let $\gamma_{(a)}$ be the set of unary literals such that for every predicate 
$P_i(x)$, $0\leq i\leq N$,
$$ 
\begin{array}{llcl}
P_i(x) &\in\gamma_{(a)} \quad&\textrm{if}&\quad \gM\models P_i(a)\\
\lnot P_i(x) &\in\gamma_{(a)} \quad&\textrm{if}&\quad \gM\not \models P_i(a).
\end{array}
$$ 
Similarly, let $\theta$ be the set of propositional literals such that 
for every proposition $p_j$, $0\leq j \leq n$,
$$
\begin{array}{llcl}
p_j &\in\theta \quad&\textrm{if}&\quad \gM\models p_j\\
\lnot p_j &\in\theta \quad&\textrm{if}&\quad \gM\not \models p_j.
\end{array}
$$

We define $\Gamma$ as $\{\gamma_{(a)} \;|\; a\in D\}$, and $\rho(c)$ as
$\gamma_{(c^{I})}$. Clearly, $\gM\models \mF_{\mC}$.
\end{proof}

\medskip

\begin{proof}[\textbf{Theorem~\ref{th:complete}: completeness of temporal
resolution}]
%\textsc{The cases when the temporal specification is unsatisfiable or there
%exists a finite path in the behaviour graph can be considered as before. (Hence,
%we assume that \TSpec{} is a satisfiable reduced specification.) The ground
%eventuality case should be considered as before.}
The proof proceeds by induction on the number of vertices in the behaviour graph
$H$ for $\TProb = \langle \uP, \iP, \sP, \eP \rangle$, which is finite.
If $H$ is empty then the set $\uP\cup\iP$ is unsatisfiable.
In this case the derivation is successfully terminated by the initial
termination rule.

Now suppose $H$ is not empty. Let $\mC$ be a vertex of $H$ which has no
successors. In this case the set $\uP\cup\mB_{\mC}$ is unsatisfiable.
Indeed, suppose $\uP\cup\{\mB_{\mC}\}$ is true in a model $\gM$. By
lemma~\ref{lem:csch},  we 
can define a colour scheme $\mC'$ such that
$\gM\models \mF_{\mC'}$.  
As $\mB_{\mC}\land\mF_{\mC'}$ is satisfiable, there exists an edge
from the vertex $\mC$ to the vertex $\mC'$ in the contradiction with the
choice of $C$ as having no successor.

The conclusion of the step
resolution rule, $\lnot \mA_{\mC}$, is added to the set $\uP$; this
implies removing the vertex $\mC$ from the behaviour graph because the
set $\{\mF_{\mC},\lnot\mA_{\mC}\}$ is not satisfiable.

\noindent Next, we check the possibility where $H$ is not empty and every
vertex $H$ has a successor. Since the problem, $\TProb$, is 
unsatisfiable, at least one condition of Theorem~\ref{th:model} is violated.
By Note~\ref{note:3to2}, it is enough to consider only two
cases of violation of the conditions of Theorem~\ref{th:model}.
\paragraph{First condition of Theorem~\ref{th:model} does not hold.}
Then, there exist a vertex $\mC_0$, predicate colour $\gamma_0$, and
eventuality $\sometime L_0(x)$ such that for every vertex $\mC'$ and predicate
colour $\gamma\in\Gamma'$,
%The negation of (\ref{eq:model1})  gives the following:
\begin{equation}\label{eq:nomodel1}
%\exists \mC\ \exists \gamma\in \Gamma\ \exists \sometime L(x)\in\eP\
%\forall\gamma\,'\in\Gamma'\ \forall \mC'\
(\mC_0,\gamma_0)\to^{+}(\mC',\gamma\,') \implies L_0(x)\notin\gamma\,'.
\end{equation}
%
%\textsc{In what follows, the proof is a modification of the proofs from
%``Simplified clausal resolution procedure for propositional linear-time
%temporal logic'' and ``Towards first-order temporal resolution''. }
%
%Let $\mC_0$, $\gamma_0$, and $\sometime L_0(x)$ be the vertex, colour and
%eventuality, respectively, determined by the existential quantifiers
%of~(\ref{eq:nomodel1}).
Let $\gI$ and $\gJ_i,i\in\gI$ be finite
nonempty sets of indexes such that $\{\mC_i\;|\; i\in \gI\}$ is
the set of \emph{all} successors of $\mC_0$ (possibly including
$\mC_0$ itself) and $\{\gamma_{i,j}\in\Gamma_i \;|\; i \in \gI,\;
j \in \gJ_i,\;\gamma_0\to^{+}\gamma_{i,j}\}$ is the set of
\emph{all} predicate colours such that there exists a run going
through  $\gamma_0$ and the colour.
%belonging to a run starting at $\gamma_0$.
(To unify notation, if $0\notin\gI$, we define $\gJ_0$ as $\{0\}$, and
$\gamma_{0,0}$ as $\gamma_0$; and if $0\in\gI$, we add the index of
$\gamma_0$ to $\gJ_0$. Therefore, $\gJ_0$ is always defined and
without loss of generality we may assume that $\gamma_{0,0} =
\gamma_0$.)

Let $\mC_{i_1}, \dots, \mC_{i_k}$ be the set of all immediate successors of
$\mC_0$. To simplify the proof, we will represent canonical
%divided merged clauses $\mA_{\mC_{i}}\implies\next\mB_{\mC_{i}}$
merged \grounded{} step clauses $\mA_{\mC_{i}}\implies\next\mB_{\mC_{i}}$
(and $\mA_{\mC_{i_l}}\implies\next\mB_{\mC_{i_l}}$) simply as
$\mA_i\implies\next\mB_i$ (and $\mA_{i_l}\implies\next\mB_{i_l}$, resp.), and
formulae $\mF_{\mC_i}$ (and $\mF_{\mC_{i_l}}$) simply as $\mF_{i}$ (and
$\mF_{i_l}$, resp.).

%For simplicity, we denote
%merged clauses $\mA_{\mC_i}\implies\mB_{\mC_i}$ as $\mA_i\implies\next\mB_i$,
%etc.

%Consider two cases depending on whether the canonical divided merged clause
Consider two cases depending on whether the canonical merged \grounded{} step clause
$\mA_0\implies\next\mB_0$ (or any of $\mA_i\implies\next\mB_i$, $i\in\gJ$)
degenerates or not.
%Consider two cases depending on whether $B_{\gamma_0}(x) = \true$ (or any of
%$B_{\gamma_{i,j}}(x) = \true$).\marginnote{be cautious}
\begin{enumerate}
\item\label{item:1}
%Let $B_{\gamma_0}(x) = \true$.
Let $\mA_0 = \mB_0 = \true$. It follows that $\uP\models\forall x\lnot
L_0(x)$. Indeed, suppose $\uP\union\{\exists x L_0(x)\}$ has a model,
%where $\Gamma'$ is one of $\Gamma_{i_1}, \dots, \Gamma_{i_k}$,
$\gM$. Then we can construct a colour scheme $\mC'$ such that
$\gM\models \mF_{\mC'}$. % and therefore
Since $\mC_{i_1}, \dots, \mC_{i_k}$ is the
set of all immediate successors of
$\mC_0$ and $\mB_0 = \true$,
it holds that there exists $j, 1\leq j \leq k$,
such that $\mC_{i_j} = \mC'$. %
%As the merged clause degenerates,
Since $B_{\gamma_0}(x) = \true$, every pair $(\gamma_0, \gamma\,')$,
where $\gamma\,'\in\Gamma'$, is suitable; hence $\lnot
L_0(x)\in\gamma\,'$ for every $\gamma\,'\in\Gamma'$, and
$\mF_{\mC'}\models\forall x \lnot L_0(x)$ leading to a
contradiction.
Therefore, $\uP\models\forall x\lnot L_0(x)$ and the eventuality
termination rule can be applied.
The same holds if any one of $\mA_i\implies\next\mB_i$ degenerates.
%\item Let neither $B_{\gamma_0}(x) = \true$ nor any $B_{\gamma_{i,j}}(x) =
%\true$. (It follows that neither $A_{\gamma_0}(x) = \true$ nor any
%$A_{\gamma_{i,j}}(x) = \true$.)
%neither $\mA_0\implies\next\mB_0$ nor any $\mA_i\implies\next\mB_i$
%degenerate.)

\item Let none of the $\mA_i\implies\next\mB_i$ degenerate. We are
going to prove that the eventuality resolution rule can be
applied. First, we have to check the side conditions for such an
application.
\begin{enumerate}
\item
$\forall x (\uP\land\mB_i\land B_{\gamma_{i,j}}(x)\implies \lnot L_0(x))$
for all $i\in \gI\cup\{0\}$, $j\in\gJ_i$.

Consider the case when
$i = j = 0$ (for other indexes the arguments are similar).

We show that
$$
\forall x (\uP\land\mB_0\land
B_{\gamma_0}(x)\implies\bigvee\limits_{l\in\{1,\dots,k\},\;\gamma\,'\in\Gamma_{i_l},\;\gamma\to\gamma\,'}F_{\gamma\,'}(x))
$$
is valid (it follows, in particular, that $\forall x (\uP\land
\mB_0\land B_{\gamma_0}(x)\implies\lnot L_0(x))$ is valid).  Suppose
$\gM$ is a model for
$$
\exists x (\uP\land\mB_0\land
B_{\gamma_0}(x)\land \bigwedge
\limits_{l\in\{1,\dots,k\},\;\gamma\,'\in\Gamma_{i_l},\;\gamma\to\gamma\,'}
\lnot F_{\gamma\,'}(x)).
$$
Then there exists a colour scheme $\mC'$ such that $\gM\models \mF_{\mC'}$.
Since $\gM\models \mB_0\land\mF_{\mC'}$, we conclude that
$\mC'$ is among $\mC_{i_1},
\dots, \mC_{i_k}$.  Note that $\gM\models
\mF_{\mC'}$ follows, in particular, $\gM\models
\forall x \bigvee\limits_{\gamma\,''\in\Gamma'} F_{\gamma\,''}(x)$ and,
hence, $\mG\models \forall x (
B_{\gamma_0}(x)\implies \bigvee\limits_{\gamma\,''\in\Gamma'}
F_{\gamma\,''}(x))$.  Together with the fact that $\gM\models\exists x
(B_{\gamma_0}(x)\land F_{\gamma\,''}(x))$ implies $\gamma_0\to\gamma\,''$, we
have
%we conclude that
$\gM\models \forall x ( B_{\gamma_0}(x)\implies
\bigvee\limits_{\gamma\,''\in\Gamma',\;\gamma_0\to\gamma\,''}
F_{\gamma\,''}(x))$.  This contradicts the choice of the structure
$\gM$.
\item $\forall x (\uP\land\mB_i\land
B_{\gamma_{i,j}}(x)\implies\bigvee
\limits_{\scriptstyle k\in\gI\cup\{0\},\;l\in\gJ_k}(\mA_k\land
A_{\gamma_{k,l}}(x)))$ for all $i\in \gI\cup\{0\}$, $j\in\gJ_i$.

Again, consider the case $i = j = 0$. Suppose
$$
\uP\land \mB_0\land \exists x (B_{\gamma_0}(x)\land\bigwedge
\limits_{\scriptstyle k\in\gI\cup\{0\},\;
l\in\gJ_k}(\lnot(\mA_k\land A_{\gamma_{k,l}}(x))))
$$
is satisfied in a structure $\gM$. Let $\mC'$ be a
colour scheme such that $\gM\models \mF_{\mC'}$.  By
arguments similar to the ones given above, there is a vertex
$\mC_{i_l}$, $1\leq l \leq k$, which is an immediate successor of
$\mC_0$, such that $\mC_{i_l} = \mC'$, and hence $\gM\models \mA'$.  It suffices to note that
$$
\gM\models \forall x (B_{\gamma_0}(x)\implies
\bigvee\limits_{\gamma\,''\in\Gamma',\; \gamma_0\to\gamma\,''}
A_{\gamma\,''}(x)) .
$$
(As in the case {2(a)} above, $\gM\models \forall x (
B_{\gamma_0}(x)\implies
\bigvee\limits_{\gamma\,''\in\Gamma',\;\gamma_0\to\gamma\,''}
F_{\gamma\,''}(x))$, and for all $\gamma\,''\in\Gamma'$, the formula
$\forall x (F_{\gamma\,''}(x)\implies A_{\gamma\,''}(x))$ is valid.)
\end{enumerate}
After applying the eventuality resolution rule we add to $\uP$ its
conclusion:
$$
\forall x\bigwedge\limits_{i\in\gI\cup\{0\},\; j\in\gJ_i} (\notl \mA_i\lor \lnot
A_{\gamma_{i,j}}(x)).
$$
Then, the vertex $\mC_0$ will be removed from the behaviour graph (recall that
$\mF_0\models\mA_0\land\exists x A_{\gamma_0}(x)$).
\end{enumerate}
\paragraph{Third condition of Theorem~\ref{th:model} does not hold.}
This case
%was already considered
%in~\cite{DF01}. We sketch here the proof.
is analogous to the previous one; we only sketch the proof.
There exist a vertex $\mC_0$ and
eventuality $\sometime l_0$ such that for every vertex $\mC'$ and predicate
colour $\gamma\in\Gamma'$,
%The negation of (\ref{eq:model2})  gives the following:
\begin{equation}\label{eq:nomodel2}
%\exists \mC\ \exists \sometime l\in\eP\ \forall \mC'\
\mC_0\to^{+}\mC' \implies l_0\notin\theta'.
\end{equation}
%Let $\mC_0$, and $l_0$ be the vertex and eventuality determined by the
%existential quantifiers of~(\ref{eq:nomodel2}).
Let $\gI$ be a finite
nonempty set of indexes, $\{\mC_i\;|\; i\in \gI\}$ be the set of all
successors of $\mC_0$ (possibly including $\mC_0$ itself). As in the
previous case, one can show that
\begin{itemize}
%\item If the canonical divided merged clause $\mA_0\implies\next\mB_0$ (or any of
\item If any of $\mA_i\implies\next\mB_i$ (where $i\in\gJ$)
  degenerates then $\uP\models\lnot l$, and the ground eventuality
  termination rule can be applied.
%\item If none of the canonical divided merged clauses degenerates then the
\item If none of the canonical merged \grounded{} step clauses
degenerate then the following conditions hold
\begin{itemize}
\item for all $i\in\gI\cup\{0\}$\qquad $\uP\cup\mB_i\models l_0$
\item for all $i\in\gI\cup\{0\}$\qquad $\uP\cup\mB_i\models
  \bigvee\limits_{j\in\gI\cup\{0\}}\mA_j$
\end{itemize}
and so the ground eventuality resolution rule can be applied.
\end{itemize}
\end{proof}
\begin{example}[example~\ref{example:graph} contd.]
We illustrate the proof of Theorem~\ref{th:complete} on the temporal
problem introduced in Example~\ref{example:graph}. The behaviour graph
of the problem is not empty; every vertex has a successor. It is not
hard to see that the first condition of Theorem~\ref{th:model} does
not hold, and, following the proof, we can choose as $\mC_0$, $\gamma_0$,
and $L_0$, for example, $\mC_1$, $\gamma_1$, and $\lnot P(x)$,
respectively.
Then for every vertex $\mC'$ and predicate colour $\gamma'\in\Gamma'$,
$$
(\mC_0,\gamma_0)\to^{+}(\mC',\gamma') \implies L_0(x)\notin\gamma'.
$$
The set of all (and all immediate) successors of $\mC_1$ is
$\{\mC_1, \mC_4\}$.  Note that the canonical full merged step clauses
corresponding to $\mC_1$ and $\mC_4$ are identical, and none of them
degenerates.  For $i \in \{1,4\}$, the loop side conditions,
$$
\forall x ((\underbrace{(l\implies \exists x P(x))}_{\uP_i}\land\underbrace{(\exists x P(x)\land\forall x P(x))}_{\mB_i}\land \underbrace{P(x)}_{B_{\gamma_1}(x)}) \implies P(x))
$$

\noindent and
$$
\begin{array}{rcl}
\forall x ((\underbrace{(l\implies \exists x
P(x))}_{\uP_i}\land\underbrace{(\exists x P(x)\land\forall x
P(x))}_{\mB_i}&\land& \underbrace{P(x)}_{B_{\gamma_1}(x)}) 
\implies\\[2.5em]
& &\bigvee\limits_{j\in\{1,4\}}(\underbrace{\exists x P(x)\land\forall x
P(x)}_{\mA_j}\land \underbrace{P(x)}_{A_{\gamma_1}(x)}))
\end{array}
$$
hold.  Therefore, we can apply the eventuality resolution rule whose
conclusion{, 
$$\forall x (\bigwedge\limits_{j\in\{1,4\}}(\lnot(\exists x P(x)
\land\forall x P(x)))\land\lnot P(x)),$$}
can be simplified to 
$
  \exists x \lnot P(x).
$
After the conclusion of the rule is added to $\uP$, vertices $\mC_1$ and $\mC_4$
and edges leading to and from them are deleted from the behaviour graph.

For the temporal problem with the new universal part, again the first condition
of Theorem~\ref{th:model} does not hold{ and}, for example, for  $\mC_0 =
\mC_3$, $\gamma_0 = \gamma_1$, and $L_0(x) = \lnot P(x)$, and for every
colour scheme $\mC'$ and every predicate colour $\gamma'\in\Gamma'$,
$$
(\mC_0,\gamma_0)\to^{+}(\mC',\gamma') \implies L_0(x)\notin\gamma'.
$$
(Note that $\gamma_2$ is never a successor of $\gamma_1$.)  The
set of all (and all immediate) successors of $\mC_3$ is $\{\mC_3,
\mC_6\}$.  The canonical full merged step clauses corresponding to
$\mC_3$ and $\mC_6$ are identical, and none of them degenerates. In a
similar way, the loop side conditions hold and
% 
% 
% (the same for $\mC_3$ and $\mC_6$)
% 
% 
% Note further that both rules
% $$
% \begin{array}{l}
% \forall x (\mA_{\mC_1} \land A_{\gamma_1}(x) \implies \next \mB_{\mC_1} \land 
%  B_{\gamma_1}(x))\\
% \forall x (\mA_{\mC_3} \land A_{\gamma_1}(x) \implies \next \mB_{\mC_3} \land 
%  B_{\gamma_1}(x))
% \end{array}
% $$
% satisfy the loop side conditions for $\sometime \lnot P(x)$. In the first case,
% however, the conclusion of the application of the eventuality resolution rule,
% does not lead to a contradiction. In the second case, 
the conclusion of the eventuality resolution rule simplifies to
$
  \forall x \lnot P(x).
$
This time, vertices $\mC_3$ and $\mC_6$ are deleted from the behaviour graph.
% together with the universal part, implies $\lnot l$ and the vertices $\mC_1$
% and $\mC_3$ are deleted from the graph.  The new universal part implies
% $\lnot l$ and the ground eventuality termination rule can be applied.

For the new problem, the third condition of Theorem~\ref{th:model} does
not hold for $\mC_0 = \mC_5$, $l_0 = l$. Then for any vertex $\mC'$,
$$
\mC_0\to^{+}\mC' \implies l_0\notin\theta'.
$$ 
As the canonical full merged step clause degenerates (and
$\uP\models\lnot l$), the ground eventuality termination rule can be
applied.

Note that if, in the beginning, instead of $\mC_1$ we selected
$\mC_3$ (or $\mC_6$) as $\mC_0$, vertices $\mC_1$, $\mC_3$, $\mC_4$, and
$\mC_6$ would be deleted after the first application of the
eventuality resolution rule.
\end{example}

\section{Extension of the Monodic Fragment}
\label{sec:extension}
In this, and the subsequent, section we adapt the resolution technique
to a number of variations of monodic \fotl{}, whose completeness
follows from the corresponding adaptation of the completeness results
given in Section~\ref{sec:proof}. We here consider an extension of
monodic temporal problems allowing an additional \emph{extended part}
$\xP$ given by a set of arbitrary \fotl{} in the language without
function symbols and with the \emph{only} temporal operator being
`$\next$'.  Since this temporal operator can be ``moved inside''
classical quantifiers, we can assume, without loss of generality, that
$\xP$ is given by a set of first-order formulae constructed from
\emph{temporal atoms} of the form $\next^i P(t_1,t_2,\dots,t_n)$,
where $P(t_1,t_2,\dots,t_n)$ is a first-order
atom\footnote{Decidability of this extension of the monodic fragment
was suggested in a private communication by M.~Zakharyaschev.}. Such
an extension permits more complex step formulae to be employed while
restricting the allowed temporal operators.

\begin{example}
\label{ex:ext-1}
A set of formulae $\XTProb$ given by
$$
\begin{array}{l}
\xP = \{ \forall x \forall y (P(x,y) \implies \next\next P(x,y)) \},\\
\iP = \{ \exists x \exists y P(x,y)\},\\
\uP = \{ \forall x \forall y (P(x,y) \implies  R(x))\}, \\
\sP = \{R(x)\implies\next R(x)\}, \\
\eP = \{\sometime \lnot R(x)\}
\end{array}
$$
\end{example}
is an example of an extended monodic problem.

We are going to show that an extended monodic temporal problem can be
translated (\emph{with a linear growth in size}) into a monodic temporal
problem while preserving satisfiability.  Essentially, we encode a few
initial states of a temporal model as a first-order formula and ensure
that this encoding is consistent with the rest of the model.

\smallskip 

%We start with the constant domain case.\marginnote{pay attention!}

\paragraph{Reduction} Let $\XTProb = \TProb\cup\xP$ be an extended monodic
temporal problem. Let $\TProb = \langle \iP, \uP, \sP, \eP
\rangle$. Let $k$ be the maximal number of nested applications of
$\next$ in $\xP$, that is, the maximal $i$ such that $\next^i
P(t_1,t_2,\dots,t_n)$ occurs in $\xP$ for some predicate symbol
$P$.  For every predicate, $P$, occurring in $\XTProb$, we
introduce $k+1$ new predicates $P^0,P^1, \ldots, P^k$ of the same
arity.  Let $\phi$ be a first-order formula in the language of $\XTProb$. We
denote by $[\phi]^i$, $0 \leq i \leq k$, the result of
substitution of all occurrences of predicates in $\phi$ with their
$i$-th counterparts; (e.g., $P(x_1, x_2)$ is replaced with
$P^i(x_1, x_2)$).

We define the monodic problem $\TProb' = \langle
\iP',\uP',\sP', \eP' \rangle$ as follows.  Let $\uP' = \uP$,
$\sP' = \sP$,  $\eP' = \eP$. As for $\iP'$, we take the following
set of formulae.
\begin{enumerate}
\item For every $\phi\in\iP$, the formula $[\phi]^0$ is in $\iP'$.
\item For every $\phi\in\uP$, the formula
$\bigwedge\limits_{i=0}^{k}[\phi]^{i}$ is in $\iP'$.
\item For every $P(x)\implies\next Q(x)\in\sP$,  the formula
   $\bigwedge\limits_{i=0}^{k-1}\left(\forall x (P^i(x)\implies
   Q^{i+1}(x)\right)$ is in $\iP'$.
\item For every $\psi\in\xP$, the formula $\psi'$, the result of
replacing all occurrences of temporal atoms
$\next^iP(\overline{t})$, $i\geq0$, in $\psi$ with
$P^{i}(\overline{t})$, is in $\iP'$.
\item For every $n$-ary predicate $P$ in the language of
$\XTProb$, the formula $\forall x_1,\dots
x_n(P(x_1,\dots,x_n)\equiv P^k(x_1,\dots,x_n))$ is in $\iP'$.

\item
No other formulae are in $\iP'$.
\end{enumerate}
\begin{example}[Example~\ref{ex:ext-1} contd.]
\label{ex:ext-2}
We give the reduction, $\TProb = \langle \uP, \iP, \sP, \eP \rangle$,
of the extended temporal problem $\XTProb$ from Example~\ref{ex:ext-1}.
The universal, step and eventuality parts of $\TProb'$ are the same 
as of $\XTProb$. The initial part, $\iP$, consists of the following
formulae
$$
\begin{array}{l}
\exists x \exists y P^0(x,y),\\
\forall x \forall y (P^0(x,y) \implies  R^0(x)),\\
\forall x \forall y (P^1(x,y) \implies  R^1(x)),\\
\forall x \forall y (P^2(x,y) \implies  R^2(x)),\\
\forall x (R^0(x)\implies R^1(x)),\\
\forall x (R^1(x)\implies R^2(x)),\\
\forall x \forall y (P^0(x,y) \implies P^2(x,y)),\\
\forall x \forall y (P(x,y)\equiv P^2(x,y)),\\
\forall x (R(x)\equiv R^2(x)).
\end{array}
$$
\end{example}

%%%%%%%%%%%%%%%%%%%%%%%%%%%%%%%%%%%%%%%%%%%%%%%%%%%%%%%%%%%%%%%%%%%%%%%%%%%%%%%
\begin{figure}[t]
\begin{pspicture}[.45](-0.5,-5.0)(4.3, 0.4)
\cnode(0,0){5pt}{x}\rput[u]{0}(0,.5){ }%
\cnode(1.0,0){5pt}{a}\rput[u]{0}(1,-0.5){$j$}\rput[u]{0}(1.0, 0.5){$\rnode{C}{ Q(\overline{\mathbf{d}})}$}%
\cnode(2,0){5pt}{b}\rput[u]{0}(2,0.5){ }%{$\varphi(x)$}%
\cnode*(3.0,0){5pt}{c}\rput[u]{0}(3.0,-0.5){$k$}%
\cnode*(4,0){3pt}{d}\rput[u]{0}(4,0.5){}%
\cnode*(5.0,0){3pt}{e}\rput[u]{0}(5,0.5){}%
\cnode*(6,0){3pt}{f}\rput[u]{0}(6,0.5){}%
\cnode*(7,0){3pt}{g}\rput[u]{0}(7,-0.5){$i+k$}\rput[u]{0}(7,.5){$\rnode{A}{P(d)}$}
\cnode*(8,0){3pt}{h}\rput[u]{0}(6,0.5){}%
\ncline{->}{x}{a}%
\ncline{->}{a}{b}%
\ncline{->}{b}{c}%
\ncline{->}{c}{d}%
\ncline{->}{d}{e}%
\ncline{->}{e}{f}%
\ncline{->}{f}{g}%
\ncline{->}{g}{h}%
\cnodeput[linecolor=white](10.0,0){i}{ }
\ncline[linestyle=dotted]{h}{i}%
%%%
\cnode(3,-1.3){5pt}{x}\rput[u]{0}(0,.5){ }%
\cnode(3.0,-2.3){5pt}{a}\rput[u]{0}(2.0,-2.3){$\rnode{D}{ Q^{j}(\overline{\mathbf{d}})}$}%
\cnode(3,-3.3){5pt}{b}\rput[u]{0}(2,0.5){ }%{$\varphi(x)$}%
\cnode*(3.0,-4.3){5pt}{c}\rput[u]{0}(3.0,-5){$0$}%
\cnode*(4,-4.3){3pt}{d}\rput[u]{0}(4,0.5){}%
\cnode*(5.0,-4.3){3pt}{e}\rput[u]{0}(5,0.5){}%
\cnode*(6,-4.3){3pt}{f}\rput[u]{0}(6,0.5){}%
\cnode*(7,-4.3){3pt}{g}\rput[u]{0}(7,-5){$i$}\rput[u]{0}(7,-3.7){$\rnode{B}{ P(d)}$}
\cnode*(8,-4.3){3pt}{h}\rput[u]{0}(6,0.5){}%
\ncline{->}{x}{a}%
\ncline{->}{a}{b}%
\ncline{->}{b}{c}%
\ncline{->}{c}{d}%
\ncline{->}{d}{e}%
\ncline{->}{e}{f}%
\ncline{->}{f}{g}%
\ncline{->}{g}{h}%
\cnodeput[linecolor=white](10.0,-4.3){i}{ }
\ncline[linestyle=dotted]{h}{i}%
\psellipse(3,-2.8)(0.6,1.9)
\nccurve[linecolor=grayb,angleA=-55,angleB=55]{<->}{A}{B}
\nccurve[linecolor=grayb,angleA=-85,angleB=130]{<->}{C}{D}
\end{pspicture}
\caption{Model transformation}\label{fig:model}
\end{figure}
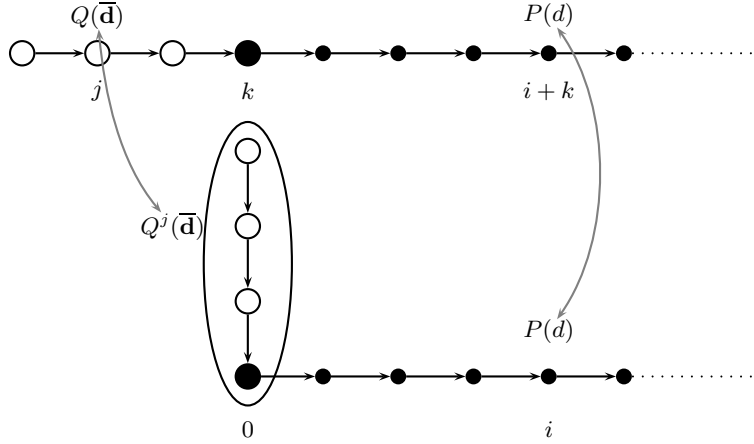
%%%%%%%%%%%%%%%%%%%%%%%%%%%%%%%%%%%%%%%%%%%%%%%%%%%%%%%%%%%%%%%%%%%%%%%%%%
\begin{theorem}[Reduction of Extended Problems]\label{th:extension}
$\XTProb$ is satisfiable if, and only if, $\TProb'$ is satisfiable.
\end{theorem}
\begin{proof}
We prove that given a model for $\XTProb$ it is possible to find a
model for $\TProb'$ and vice versa. The transformation of models is 
depicted in Fig.~\ref{fig:model}.

First, consider a model
$\gM=\langle D, I \rangle$ for $\XTProb$ and construct a model $\gM' =
\langle D, I' \rangle$ as follows.  The interpretation of constants in
the language of $\XTProb$ in $\gM'$ is the same as in $\gM$ (recall
that constants are \emph{rigid}).

 For every $n$-ary predicate $P$ in the language of
$\XTProb$ (in the initial signature), every $n$-tuple
$(d_1,\dots,d_n)\in D$, and every $i\geq0$, we define
$$
\gM'_i\models P(d_1,\dots, d_n)\quad
\text{ iff }\quad
\gM_{i+k}\models P(d_1,\dots, d_n).
$$
For every $n$-ary predicate $P^i$, $0 \leq i \leq k$, in the
extension of the initial language (that is, in the language of
$\TProb'$ but not in the language of $\XTProb$)
  we define
$$
\gM'_0\models P^i(d_1,\dots, d_n) \quad\text{ iff }\quad \gM_i\models
P(d_1,\dots, d_n),
$$
and $P^i$ is false in $\gM'$ for all other tuples and moments of
time\footnote{Note that all new predicates occur only in  $\iP'$.}.
This definition is consistent with formulae from part 5 of $\iP'$;
therefore $\gM'$ is defined correctly.

Since truth values of all predicates from $\TProb$ are not changed
but ``shifted'', clearly, $\gM'\models \uP$ and $\gM'\models\sP$.
Since all our eventualities are unconditional, that is, are of the
form $\always\sometime l$ and $\always\forall x \sometime L(x)$,
the truth value of $L(x)$ in the first $k+1$ states of $\gM$ does not
affect the truth value of $\eP'$ in $\gM'$; so $\gM'\models\eP'$. The
fact that $\gM'\models\iP'$ can be established by considering step
by step the definition of $\iP'$. Indeed:
\begin{enumerate}
\item Let a formula  $[\phi]^0$ be in $\iP'$, where $\phi\in\iP$.
Then $\gM'_0 \models [\phi]^0$ because for every predicates $P$
and $P^0$,
$$
\gM_0\models P(d_1,\dots,d_n)\quad \text { iff }\quad \gM'_0\models
P^0(d_1,\dots,d_n)
$$
holds and $\gM_0\models \phi$.

\item Let a formula  $[\phi]^i$, $0 \leq i \leq k$, be in $\iP'$,
where $\phi\in\uP$. Then $\gM'_0 \models [\phi]^i$ because for
all predicates $P$ and $P^i$,
$$
\gM_i\models P(d_1,\dots,d_n) \quad\text { iff }\quad \gM'_0\models
P^i(d_1,\dots,d_n)
$$
holds and $\gM_i\models \phi$.

\item Let $\gM'_0\models P^i(d_1,\dots,d_n)$, $0 \leq i \leq k$.
Then $\gM_i\models P(d_1,\dots,d_n)$, and because of
$P(x)\implies\next Q(x)\in\sP$, we have $\gM_{i+1}\models
Q(d_1,\dots,d_n)$. It follows $\gM'_0\models
Q^{i+1}(d_1,\dots,d_n)$.

\item Let $\psi\in\xP$, that is, $\gM_0\models \psi$. For every
subformula $\next^iP(d_1,\ldots, d_n)$ of $\psi$, $\gM_0\models
\next^iP(d_1,\ldots, d_n)$ holds if, and only if, $\gM'_0\models
P^i(d_1,\ldots, d_n)$. So, $\gM'_0\models \psi'$.

\item In accordance with the definition of $\gM'$, 
$\gM'_0\models P(d_1,\ldots, d_n)$ if, and only if,
$\gM_k\models P(d_1,\ldots, d_n)$ if, and only if,
$\gM'_0\models P^k(d_1,\ldots, d_n)$.
\end{enumerate}
Let $\gM'$ be a model for $\TProb'$. We construct a model $\gM$ for
$\XTProb$.  The interpretation of constants in the language of
$\XTProb$ in $\gM$ is the same as in $\gM'$.  For every $n$-ary
predicate $P$ in the language of $\XTProb$ and every $n$-tuple
$(d_1,\dots,d_n)\in D$ we define for every $i\geq k$
$$
\gM_i\models P(d_1,\dots,d_n)\quad
\text { iff }\quad
\gM'_{i-k}\models P(d_1,\dots,d_n),
$$
and for every $i$ such that  $0\leq i \leq k$
$$
\gM_i\models P(d_1,\dots,d_n) \quad\text { iff }\quad \gM'_0\models
P^i(d_1,\dots,d_n).
$$
Note that $\gM'_0\models\iP'$ and, in particular, formulae from
part 5 of $\iP'$; therefore, $\gM$ is defined correctly. Indeed,
in the case $i=k$ we obtain
$$
\gM'_0\models P(d_1,\dots,d_n)\quad \text { iff }\quad \gM'_0\models
P^k(d_1,\dots,d_n).
$$
Evidently, for $i\geq k$, $\gM_i\models \uP$ and $\gM_i\models
\sP$. Again, since our eventualities are unconditional, evaluation of
$\eP$ does not depend on a finite number of initial states, and
$\gM\models\eP$. It is enough to show that $\gM_i\models \uP$ and
$\gM_i\models \sP$ for $i \in [0,(k-1)]$,
%($0\leq i\leq (k-1)$),
%$i:0\leq i\leq (k-1)$,
and $\gM_0\models
\iP$. Again, this can be done by analysing the definition of $\iP'$.
\smallskip

\noindent The first claim, $\gM_i\models \uP$, follows from item 2
of the definition of $\iP'$, from the relation
$$ \gM_i\models
P(d_1,\dots,d_n) \quad\text { iff }\quad \gM'_0\models P^i(d_1,\dots,d_n)
$$
and the fact that $\gM'_0\models [\phi]^i$ for every $\phi \in \uP$,\
$0\leq i\leq k$.
\smallskip

\noindent The second claim, $\gM_i\models \sP$,
%\ $0\leq i\leq (k-1)$,
follows from item 3
of the definition of $\iP'$ and from the relation
$$
\gM_i\models P(d_1,\dots,d_n)\quad \text { iff }\quad \gM'_0\models
P^i(d_1,\dots,d_n).
$$
The last claim, $\gM_0\models \iP$, follows immediately
from item 1 of the definition of $\iP'$ and from the relation
$$
\gM_0\models P(d_1,\dots,d_n)\quad \text { iff }\quad \gM'_0\models
P^0(d_1,\dots,d_n)
$$
given above.
\end{proof}

\section{Grounding Temporal Problems}
%\marginnote{Is it a good name?}
\label{sec:properties}
%
% (Up to satisfiability we can restrict ourselves only by the 
% future-time operators $\always$, `always', $\next$, `next-time' and
% $\sometime$, `sometime', or `eventually'.)
%
In this section we adapt the core temporal resolution calculus given
in Section~\ref{sec:ccalc} to a variation of monodic \fotl{} where
sub-parts of the temporal problem are \emph{grounded}. Not only does
this characterise an important class of formulae, but this variation
admits simplified clausal resolution techniques (in particular,
simplified DSNF).
\begin{definition}[Groundedness]
A temporal problem $\TProb$ is called {\em grounded} if all the step
clauses and the eventuality clauses of $\TProb$ are
ground. Correspondingly, a temporal monodic formula is called grounded
if it can be translated to a grounded temporal problem.  A temporal
problem $\TProb$ is called a {\em ground eventuality} problem if all
the eventualities of $\TProb$ are ground.  A temporal problem $\TProb$
is called a {\em ground next-time} problem if all the step clauses of
$\TProb$ are ground.
%% Correspondingly, a temporal monodic formula
%% is called a ground next-time formula if it is translated to a
%% ground next-time temporal problem.
\end{definition}
%%
%Formulae in which non-ground
%formulae are only allowed under the next-time operator are called
%\emph{ground eventuality} monodic formulae (this terminology came
%from DSNF and will be explained later).  Formulae in which
%non-ground  expressions are only allowed under the sometime operator
%are called \emph{ground next-time} formulae.
%
%In
%Section~\ref{sec:properties} we show that
%any ground eventuality formula can be reduced (with an exponential
%growth in size) to a satisfiability equivalent near-propositional formula;
%any ground next-time formula can be reduced (with a linear growth in size)
%to a satisfiability equivalent near-propositional formula;
%and that over finite domains (a characteristic case in applications)
%any monodic formula can be reduced (with a linear growth in size) to
%a satisfiability equivalent ground eventuality formula.
%
If $\TProb$ is a ground eventuality problem then only the ground
versions of the eventuality resolution and eventuality termination
rules are needed. 
\begin{theorem}[Reducing a Ground Eventuality Problem]\label{th:1}
Every ground eventuality monodic temporal problem can be
reduced to a satisfiability equivalent grounded
monodic problem with an exponential growth in size of the given
problem.
\end{theorem}
\begin{proof}
Note that the ground eventuality resolution rule,
step resolution rule, and initial termination rule operate on merged
derived step clauses. So, if instead of original step clauses we
consider step clauses given by formulae $(\ref{eq:const})$,
$(\ref{eq:exists})$, and $(\ref{eq:forall})$ (and strictly speaking,
rename by propositions closed first-order formulae in the right- and
left-hand sides), we obtain a satisfiability equivalent grounded
temporal problem.
\end{proof}
%The reduction is exponential in size of the original problem.
\begin{example}
Consider an unsatisfiable formula
$$
\sometime \exists x (P(x)\land \next\lnot P(x))\land\always(P(x)\implies\next
P(x)).  
$$
In DSNF we have (note that $\iP$ is empty throughout),
$$
\begin{array}{ll}
\sP = \left\{
\begin{array}{l}
P(x)\implies\next P(x)\\
Q(x)\implies\next\lnot P(x)
\end{array}
\right\},\qquad &
\begin{array}{l}
\uP = \emptyset,\\[1ex]
\eP = \{\sometime \exists x (P(x)\land Q(x))\}.
\end{array}
\end{array}
$$
According to our reduction, this problem is satisfiability equivalent to the  following
$$
\begin{array}{l}
\uP = \emptyset,\\
\sP = \left\{
\begin{array}{l}
\exists x P(x)\implies\next \exists x P(x)\\
\forall x P(x)\implies\next \forall x P(x)\\
\exists x Q(x)\implies\next \exists x \lnot P(x)\\
\forall x Q(x)\implies\next \forall x \lnot P(x)\\
\exists x (P(x)\land Q(x)) \implies\next \exists x (P(x)\land\lnot P(x))\\
\forall x (P(x)\lor Q(x)) \implies\next \forall x (P(x)\lor\lnot P(x))\\
\end{array}
\right\},\qquad \\
\eP = \{\sometime \exists x (P(x)\land Q(x))\}.
\end{array}
$$
The last step clause is a tautology which can be eliminated immediately,
the next to last can be moved to the universal part by an application
of step resolution. .
$$
\begin{array}{l}
\uP = \{\forall x (\lnot P(x)\lor\lnot Q(x))\},\\
\sP = \left\{
\begin{array}{l}
\exists x P(x)\implies\next \exists x P(x)\\
\forall x P(x)\implies\next \forall x P(x)\\
\exists x Q(x)\implies\next \exists x \lnot P(x)\\
\forall x Q(x)\implies\next \forall x \lnot P(x)
\end{array}
\right\},\qquad \\
\eP = \{\sometime \exists x (P(x)\land Q(x))\}.
\end{array}
$$
Now the ground eventuality termination rule can be applied.
\hfill\qed
\end{example}
Together with Theorem~\ref{th:1} the following theorem shows that for
any problem $\TProb$, if either all the step clauses are ground or all
the eventuality clauses are ground, then it can be reduced to a grounded
problem.

\begin{theorem}[Reducing a Ground Next-time Problem]\label{cor:2}
\mbox{}\\ Let $\TProb = \langle \iP, \uP, \sP, \eP\rangle$ be a temporal
 problem such that all step rules of $\TProb$ are ground.
Let $\eP^{\exists}$ be obtained from $\eP$
as follows: every eventuality clause of the form $\sometimes
L(x)$ (in the meaning of $\forall x \sometimes
L(x)$)
is replaced with its  \emph{ground} consequence $\exists x
\sometimes L(x)$ (equivalent to $\sometimes \exists x L(x))$. Let
$\TProb' = \langle \iP, \uP, \sP, \eP^{\exists} \cup
\{\sometime L(c)\;|\; \sometime L(x)\in\eP,\ c\in \const(\TProb) \}\rangle$.
%\{\sometime
%L(c)\;|\; L(x)\in\eP, c\in\mL \}\rangle$.
Then $\TProb$ is satisfiable if
and only if $\TProb'$ is satisfiable.
\end{theorem}
\begin{proof}(Sketch)
%Consider now a constant-free
%temporal problem such that all step
%clauses in $\TProb$ are ground.
Evidently, if $\TProb'$ is unsatisfiable, then $\TProb$ is unsatisfiable. Suppose
now $\TProb$ is unsatisfiable, then there exists a successfully terminating
temporal resolution derivation from $\TProb^c$. Note that the added eventualities
of the form $\sometime L(c)$ exactly correspond to the eventualities
added by reduction to constant-flooded form.

Suppose
the eventuality resolution rule is applied to
a non-ground eventuality $\forall x \sometimes L(x)$.
%is involved in the temporal resolution derivation
%of a contradiction (that is,  the eventuality resolution rule or the
%eventuality termination rule is applied to this eventuality). In such case the
The validity
%google says that validity occurs 2 700 000 times, validness -- 800 times
of the side conditions implies the validity of the formula
\begin{equation}
\label{nt1:eq1}%
 \always \forall x (\uP \land
\bigvee\limits_{j=1}^n \mA_i \implies \next \always \lnot L(x))
\end{equation}
for a set $\{ \mA_i \imp \next \mB_i, 1 \leq i \leq \}$ of
%\emph{divided} merged step rules.
\emph {ground} merged {\grounded} step rules.
%compounded of propositional step rules of $\TProb$.
%In
%applying the temporal resolution rule
(\ref{nt1:eq1}) is resolved with the formula $\always \forall
x \sometimes L(x)$ giving the conclusion $(\bigwedge\limits_{i=1}^n
\notl \mA_i)$. However (\ref{nt1:eq1}) or, equivalent to
(\ref{nt1:eq1}),
\begin{equation}
\label{nt1:eq2}%
 \always (\uP \land
\bigvee\limits_{j=1}^n \mA_i \implies \next \always \forall x
\lnot L(x))
\end{equation}
can be resolved with a ``weaker'' formula $\always \exists x
\sometimes L(x)$ giving the same result.
%So we can conclude that
%replacing non-ground eventualities of the form $ \forall x
%\sometimes L(x)$ with ground eventualities $ \exists x
%\sometimes L(x)$ (equivalent to $ \sometimes \exists x L(x))$
%does not affect satisfiability.

If the eventuality termination rule is applied to $\forall x \sometimes L(x)$,
its side condition, $\uP\models\forall x\lnot L(x)$, equally contradicts to
the ground eventuality $\exists x \sometimes L(x)$.
So, we can conclude that
replacing non-ground eventualities of the form $ \forall x
\sometimes L(x)$ with ground eventualities $ \exists x
\sometimes L(x)$ (equivalent to $ \sometimes \exists x L(x))$
does not affect (un)satisfiability.
%suggests the following two lemmas, whose
%proofs can be found in Appendix~\ref{sec:proofs}.
\end{proof}
\begin{example}
$$
\begin{array}{ll}
\iP = \{l\},\qquad & \uP = \{\forall x (l\implies Q(x))\},\\
\sP = \{
l\implies\next l
\},\qquad &
\eP = \{\sometime \lnot Q(x)\}.
\end{array}
$$ Evidently, the initial, universal, and step parts imply
$\always\forall x Q(x)$ which also contradicts to $\always \forall x
\sometime \lnot Q(x)$ and $\always \exists x \sometime \lnot Q(x)$.
\hfill\qed
\end{example}

\section{Decidability by Temporal Resolution}\label{sec:classes}
%
%We explore here decidability consequences of temporal resolution.
%Note that side conditions of all rules of temporal resolution
%Theorem~\ref{th:complete} produces a
Temporal resolution provides a decision procedure for a class of
monodic temporal formulae provided that there exists a first-order
decision procedure for side conditions of all inference rules.  Direct
examination of the side conditions shows that we are interested in the
satisfiability of the conjunction of the (current) universal part and sets
of monadic formulae built from predicate symbols which occur in the
temporal part.
%For example, if we take as a side condition of the
%eventuality resolution rule w.r.t. (a current universal part)
%$\uP$ a formula
%\[\forall x (\uP  \land \mB_i \land B_i(x) \implies
%\bigvee\limits_{j=1}^n (\mA_j\land A_j(x))
%\]
%then the validity of this formula is equivalent to satisfiability
%of a set
%\[ \uP \cup \mB_i \cup
%\exists x (B_i(x) \andl \bigwedge \limits_{j=1}^n( \notl \mA_j
%\orl \notl A_j(x))
%\]
At the same time, the current universal part of a derivation is
obtained by extending the initially given universal part by
monadic formulae from the conclusions of
the inference rules.
%step and temporal
%resolution inferences.
So, after imposing appropriate restrictions on the form of the
universal part of a given temporal problem, we can guarantee its
decidability (the addition of monadic formulae usually does not affect
decidability).
%\cite{FLHT01}.

%temporal resolution rules. Direct examination of the side
%conditions shows that we are interested in satisfiability of the
%those formulae that occur in  $\uP$-part of a given problem
%together with conjunction of closed formulae each of which is
%either universally or existentially quantified or contains
%constants. In this aspect, our approach differs from~\cite{HWZ00}
%where the underlying first-order fragment needed to be closed
%under negation (which is not the case for the fragments that we
%consider here).

%
To reflect our ``rename and unwind'' transformation to the normal
form, we define decidable fragments in terms of
\emph{surrogates}~\cite{HWZ00}. Let us reserve for every formula
$\phi$, whose main connective is a temporal operator, a unary predicate
$P_\phi(x)$, and for every sentence $\psi$, whose main connective is a
temporal operator, a propositional variable $p_\psi$. $P_\phi(x)$ and
$p_\psi$ are called surrogates. Given a monodic temporal formula
$\phi$, we denote by $\overline{\phi}$ the formula that results from
$\phi$ by replacing all of its subformulae whose main connective is a
temporal operator and which is not within a scope of another temporal
operator with their surrogates.

Such an approach allows us to define decidable monodic classes based
on the properties of surrogates analogously to the classical first-order
decision problem~\cite{BGG97}. Note however, that it is necessary to
take into consideration occurrences of temporal operators as the
following example shows.
%\subsection{Prefix classes}
%Unlike classical first-order decision problem~\cite{BGG97}
%deals with collections of
%prenex first-order formulae characterised by the form of the quantifier prefix
%and/or vocabulary;
%where any first-order formula can be transformed into an
%equivalent one in prenex normal form,
%as for temporal formulae,
%reduction of a monodic formula to DSNF can change the shape of the
%quantifier prefix.\marginnote{Change this sentence.}

\begin{example}
The first-order formula $\exists x \forall y \forall z \exists u
\Phi(x,y,z,u)$, where $\Phi$ is quantifier free, belongs to the
classical decidable fragment \EAAE. Let us consider the temporal
formula $\exists x \always \sometime \forall y \forall z \exists u
\Phi(x,y,z,u)$ with the same $\Phi$. It is not hard to see that after
the translation into DSNF (see Example~\ref{ex:dsnf2}), the first
formula from $\uP$ does not belong to \EAAE any more.  (Formally, it
belongs to the undecidable Sur\'anyi class $\forall^3\exists$.)
\hfill\qed
\end{example}

%\cite{Kontchakovetal02}
%
%Note that in absence of function symbols, there exist only three maximal
%classical decidable classes, $\exists^*\forall^*$, \EAAE, and monadic fragment.
\noindent The following definition takes into account the
considerations above.
\begin{definition}[Temporalisation by Renaming]
Let $\gC$ be a class of first-order formulae.  Let $\phi$ be a
monodic temporal formula in Negation Normal Form (that is, the
only boolean connectives are conjunction, disjunction and
negation, and  negations are only applied to atoms). We say that
$\phi$ belongs to the class $\Tren\gC$
%\TEAAE
if
\begin{enumerate}
\item $\overline{\phi}$ belongs to
$\gC$
%\EAAE
and \item for every subformula of the form $\mT\psi$, where $\mT$
is a temporal operator (or of the form $\psi_1\mT\psi_2$ if $\mT$
is binary), either $\overline{\psi}$ is a closed formula belonging
to $\gC$ or the formula $\forall x (P(x) \implies
\overline{\psi})$, where $P$ is a new unary predicate symbol,
 belongs to $\gC$ (analogous conditions for $\psi_1$, $\psi_2$).
%$\forall x \overline{\psi}$ (or both $\forall x \overline{\psi_1}$ and
%$\forall x \overline{\psi_2}$) is either a closed formula
%belonging to \EAAE or belongs to \AAE.
\end{enumerate}
\end{definition}
Note that the  formulae indicated in the first and second items of
the definition exactly match the shape of the formulae
contributing to $\uP$ when we reduce a temporal formula to the
normal form by renaming the complex expressions and replacing
temporal operators by their fixpoint definitions.
\begin{theorem}[Decidability by Temporal Resolution]
\label{thm:decidability}%
 Let $\gC$ be a decidable class of
first-order formulae which does not contain equality and
functional symbols, but possibly contains constants, such that
\begin{itemize}
\item $\gC$ is closed under conjunction;
\item $\gC$ contains monadic formulae.
%\item for every quantifier-free formula $\Phi$ with at most one free variable,
%formulae $\forall x \Phi$ and $\exists x \Phi$ belong to $\gC$.
\end{itemize}
Then $\Tren\gC$ is decidable.
\end{theorem}
\begin{proof}
After reduction to DSNF, all formulae from $\uP$ belong
to $\gC$. The (monadic) formulae from side conditions and the (monadic)
formulae generated by temporal resolution rules belong to $\gC$.
%because any
%bounded monadic formula is transformed into an equivalent
%conjunction of formulae of the form $\forall x \Phi$ or $\exists x
%\Phi$ where $\Phi$ is a monadic formula with at most one free
%variable.
Theorem~\ref{th:complete}% and \ref{th:completeE} 
gives the decision procedure.
\end{proof}
Theorem~\ref{thm:decidability} provides the possibility of using
temporal resolution to confirm decidability of all temporal monodic
classes listed in \cite{HWZ00,WZ:APAL:AxMono}: \emph{monadic,
two-variable, fluted, guarded and loosely guarded}. Moreover,
combining the constructions from~\cite{Hodkinson00} and the
%(first-order)
saturation-based decision procedure for the guarded fragment with
equality~\cite{GN99}, it is possible to build a temporal resolution
decision procedure for the monodic guarded and loosely guarded fragments
with equality~\cite{DFK03LPAR}.
%(but without functional symbols).
%\footnote{aaa}.

In addition, using the above theorem, we also obtain decidability of
some monodic \emph{prefix-like} classes.

\begin{corollary}[$\Tren\EAAE$, temporalised G\"{o}del class]
The class $\Tren\EAAE$ is decidable
\end{corollary}
\begin{proof}
Every monadic formula can be reduced, in a satisfiability
equivalence preserving way, to a conjunction of formulae of the
form $\forall x (l_{1}\lor\dots\lor l_{p}\lor
L_{1}(x)\lor\dots\lor L_{q}(x))$, $p,q\geq0$ or $\exists x
(L_{1}(x)\land\dots\land L_{r}(x))$, $r\geq0$, where $l_{j}$ are
ground literals and $L_{j}(x)$ are non-ground literals. Obviously,
every
conjunct is in \EAAE. Satisfiability of a conjunction of formulae
belonging to \EAAE is decidable, e.g. by the resolution-based
technique (see clause set class $\mathcal{S}^+$ in
~\cite{FLHT01}).
\end{proof}
\begin{corollary}[$\Tren K$, temporalised Maslov class]
The class $\Tren K$ is decidable (where $K$ is the Maslov class).
\end{corollary}
\begin{proof}
Again, monadic formulae can be rewritten as a conjunction of Maslov formulae;
satisfiability of a conjunction of Maslov formulae is decidable as shown
in~\cite{HS99a}.
\end{proof}

\newcommand{\lf}{\mathbf}
\section{Loop Search Algorithm}
\label{sec:loopsearch}
The notion of a full merged step clause given in Section~\ref{sec:proof} 
is quite involved and the
search for appropriate merging of simpler clauses is computationally
hard. Finding \emph{sets} of such full merged clauses needed for the
temporal resolution rule is even more difficult.
%Note that 
%only the left-hand sides of full merged step clauses contribute to the 
%result of the rule application.  
In 
%this section 
Fig.~\ref{fig:bfs}
we present a search algorithm that finds a \emph{loop
formula} (cf. page~\pageref{page:loop}) --- a disjunction of the
left-hand sides of full merged step clauses that together with an
eventuality literal form the premises for the temporal resolution
rule. The algorithm is based on Dixon's loop search algorithm for
the propositional case~\cite{Dixon96}. For simplicity, in what
follows we consider non-ground eventualities only.
The algorithm and the proof of its properties for the ground case can 
be obtained by considering merged \grounded{} step clauses instead of the 
general case and by deleting the parameter ``$x$'' and quantifiers.
%from the text below.
%%%%%%%%%%%%%%%%%%%%%%%%%%%%%%%%%%%%%%%%%%%%%%%%%%%%%%%%%%%%%%%%%%%%%%%%%%
\begin{figure*}[t]
\fbox{
\begin{minipage}{0.97\textwidth}
\small
%\paragraph{Breadth First Search algorithm.}
\begin{description}
\item[Input] A temporal problem $\TProb$ and an eventuality clause 
$\sometime L(x)\in\eP$.
\item[Output] A formula $H(x)$ with at most one free variable.
\item[Method:]~\mbox{}\hspace{-1.5em}
\begin{minipage}[t]{0.92\textwidth}
\begin{enumerate}\itemsep=0pt
\item Let $H_0(x) = \true$; $N_0 = \emptyset$; $i = 0$.
\item\label{label:loop} Let $N_{i+1} = \{\forall x (\mA^{(i+1)}_j(x)\implies\next\mB^{(i+1)}_j(x))\}_{j=1}^{k}$ be the set
 of \emph{all} full merged step clauses such that for every
 $j\in\{1\dots k\}$, $\forall x (\uP\land\mB^{(i+1)}_j(x) \implies
 (\lnot L(x)\land H_i(x)))$ holds.  (The set $N_{i+1}$ possibly
 includes the degenerate clause $\true\implies\next\true$ in the case
 $\uP\models\forall x (\lnot L(x)\land H_i(x))$.) 
\item\label{label:exit1} If $N_{i+1} = \emptyset$, 
 % let $H_{i+1}(x) = \false$; 
 return $\false$;
 else let 
 $H_{i+1}(x) = \bigvee\limits_{j=1}^{k}(\mA^{(i+1)}_j(x))$.
%\item\label{label:exit1} If $N_{i+1} = \emptyset$ return $\false$.
\item\label{label:exit2} If $\forall x (H_i(x)\implies H_{i+1}(x))$ return $H_{i+1}(x)$.
\item $i = i+1$; goto~\ref{label:loop}.
\end{enumerate}
\end{minipage}
\end{description}
\end{minipage}
} 
\caption{Breadth First Search algorithm.\label{fig:bfs}}
\end{figure*}%
%%%%%%%%%%%%%%%%%%%%%%%%%%%%%%%%%%%%%%%%%%%%%%%%%%%%%%%%%%%%%%%%%%%%%%%%%%
We are going to show now that the algorithm terminates
(Lemma~\ref{lem:termination}), its output is a loop formula 
(lemmas~\ref{lem:loopformula} and \ref{lem:implies}), and temporal
resolution is complete if we consider only the loops
generated by the algorithm (Theorem~\ref{th:relcomp}).

\begin{lemma}\label{lem:raw}
The formulae $H_i(x)$, $i\geq0$, constructed by the BFS algorithm, satisfy
the following property: $\forall x (H_{i+1}(x) \implies H_{i}(x))$.
\end{lemma}
\begin{proof}
By induction. In the base case $i=0$, we have $H_0(x)\equiv\true$ and,
obviously, $\forall x (H_1(x)\implies\true)$. The induction hypothesis is that
$\forall x (H_{i}(x) \implies H_{i-1}(x))$. In the induction step, let 
$N_{i+1}\neq\emptyset$ (otherwise, $H_{i+1}(x)\equiv\false$ and, evidently,
$\forall x (\false\implies H_i(x))$ holds). Let $N_{i+1} = \{\forall x
(\mA^{(i+1)}_j(x)\implies\next\mB^{(i+1)}_j(x))\}_{j=1}^{k}$. For every
$j\in\{1\dots k\}$ we have $\forall x (\mB^{(i+1)}_j(x)\implies(\lnot L(x)\land
H_{i}(x)))$. By the induction hypothesis, 
%$\forall x (\mB_j(x)\implies(\lnot L(x)\land H_{i-1}(x)))$, 
$\forall x (H_{i}(x)\implies H_{i-1}(x))$ and, therefore, $\forall x
(\mB^{(i+1)}_j(x)\implies(\lnot L(x)\land H_{i-1}(x)))$, that is,
$N_{i+1}\subset N_{i}$. It follows that $\forall x (H_{i+1}(x)\implies
H_{i}(x))$.
\end{proof}
\begin{lemma}\label{lem:termination} The BFS algorithm terminates.
\end{lemma}
\begin{proof}
There are only finitely many different $H_i(x)$. Therefore, either
there exists $k$ such that $H_k(x) \equiv \false$ and the algorithm terminates
by step~\ref{label:exit1}, or there exist
$l,m: l<m$ such that $\forall x (H_l(x)\equiv H_m(x))$. 
{In} the
latter case, by Lemma~\ref{lem:raw}, we have $\forall x
(H_{m-1}(x)\implies H_l(x))$, that is $\forall x (H_{m-1}(x)\implies
H_{m}(x))$.
By step~\ref{label:exit2}, the algorithm
terminates.
\end{proof}
\begin{lemma}\label{lem:loopformula}
Let $H(x)$ be a formula produced by the BFS algorithm. Then
$\forall x (\uP\land H(x)\implies\next\always \lnot L(x))$.
\end{lemma}
\begin{proof}
If $H(x) = \false$, the lemma holds. Otherwise, consider the last computed set
$N_{i+1}$ (that is, $H(x) = H_{i+1}(x)$). Let $N_{i+1} = \{\forall x
(\mA^{(i+1)}_j(x)\implies\next\mB^{(i+1)}_j(x))\}_{j=1}^{k}$.
Note that for all $j\in\{1\dots k\}$, it holds
 $\forall x (\uP\land\mB^{(i+1)}_j(x) \implies \lnot L(x))$ and, since 
 $\forall x (H_i(x)\implies H_{i+1}(x))$, we also have
 $\forall x (\uP\land\mB^{(i+1)}_j(x) \implies H_{i+1}(x))$, that is, $N_{i+1}$
 is a loop and $H_{i+1}(x)$ is its loop formula.
\end{proof}
\begin{lemma}\label{lem:implies}
Let $\TProb$ be a monodic temporal problem,
$\mL$ be a loop in $\sometime L(x)\in\eP$, and $\lf{L}(x)$ be its loop formula.
Then for the formula $H(x)$, produced by the BFS algorithm on % $\TSpec$ and
$\sometime L(x)$, the following holds: $\forall x (\lf{L}(x)\implies H(x))$.
\end{lemma}
\begin{proof}
We show by induction that for all sets of full merged step clauses $N_{i+1}$, 
constructed by the algorithm,
%holds $\forall x (\lf{L}(x)\implies H_{i+1}(x))$. Indeed, 
$\mL\subset N_{i+1}$. In the base case $i=0$, $H_0(x)\equiv\true$ and
for every full merged step clause $\forall x(\mA(x)\implies\mB(x))\in\mL$, 
we have 
$\forall x (\uP\land\mB(x)\implies (\lnot L(x)\land\true))$; therefore, 
$\mL\subset N_1$.% and $\forall x(\lf{L}(x)\implies H_1(x))$.

Our induction hypothesis is that $\mL\subset N_i$, that is, $N_i = \mL \cup
N_i'$.  Then $H_i(x) = \lf{L}(x)\lor H_i'(x)$. Let $\forall
x(\mA(x)\implies\mB(x))$ 
be any full merged step clause from $\mL$. By the definition of a loop,
$\forall x (\uP\land \mB(x)\implies (\lnot L(x)\land\lf{L}(x)))$, hence,
$\forall x (\uP\land \mB(x)\implies ((\lnot L(x)\land\lf{L}(x))\lor (\lnot
L(x)\land H_i'(x))))$, that is, $\forall x (\uP\land\mB(x)\implies (\lnot
L(x)\land H_i(x)))$.  
Since the set $N_{i+1}$ consists of all full merged step clauses, 
$\forall x (\mA^{(i+1)}_j(x)\implies\next\mB^{(i+1)}_j(x))$, such that
$\forall x(\uP\land\mB^{(i+1)}_j(x) \implies (\lnot L(x)\land H_i(x)))$ holds,
we have $\forall x(\mA(x)\implies\mB(x))\in N_{i+1}$. As $\forall
x(\mA(x)\implies\mB(x))$ is an arbitrary full merged step clause from $\mL$,
it means that $\mL\subset N_{i+1}$.

It follows that $\forall x (\lf{L}(x)\implies H(x))$.
\end{proof}
The proof of the completeness theorem goes by showing that there
exists an eventuality $\sometime L(x)\in\eP$ and a loop $\mL=\{\forall
x (\mA_i(x)\implies\next\mB_i(x))\}_{i=1}^{k}$ such that the application
of the eventuality resolution rule to $\sometime L(x)$ and $\mL$ leads
to the deletion of some vertices from the eventuality graph.
A vertex $\mC$ is deleted from the graph if the categorical formula,
$\mF_{\mC}$, together with the universal part, $\uP$, is
satisfiable, but $\mF_{\mC}\land\forall x \lnot \bigvee_{j=1}^{k}\mA_j(x)\land\uP$ is
unsatisfiable.
\begin{theorem}[Relative Completness]\label{th:relcomp}
Temporal resolution is complete if we restrict ourselves to loops found 
by the BFS algorithm.
\end{theorem}
\begin{proof}
Let $H(x)$ be the output of the BFS algorithm, let
$\lf{L}(x)\eqbydef\bigvee_{j=1}^{k}\mA_j(x)$. By
Lemma~\ref{lem:implies}, $\forall x (L(x)\implies H(x))$ holds; therefore,
$H(x)$ is not $\false$. 
From the proof of Lemma~\ref{lem:loopformula} it follows
that the
last computed set $N_{i+1}$ (that is, $H(x) = H_{i+1}(x)$) is a loop in 
$\sometime L(x)$ and $H(x)$ is its loop formula. 
Since $\forall x (\lf{L}(x)\implies H(x))$, 
the formula $\mF_{\mC}\land\forall x \lnot H(x)\land\uP$ is unsatisfiable as
well and the application of the eventuality resolution rule to $\sometime L(x)$
and $N_{i+1}$ leads to deletion of at least the same vertices from the eventuality 
graph.
\end{proof}
\begin{note}
The need to include \emph{all} full merged step clauses satisfying some particular
conditions into $N_{i+1}$ might lead to quite extensive computations. Note
however that due to the trivial fact that if $\forall x(A(x)\implies B(x))$ 
then $\forall x ((A(x) \lor B(x))\equiv B(x))$,
we can restrict the choice to only those full merged step clauses whose left-hand
sides do not imply the left-hand side of any other clause in $N_{i+1}$ yielding a
formula $H_{i+1}'(x)$ equivalent
to the original formula $H_{i+1}(x)$.
% 
% Alternatively, we can perform \emph{simplification} of 
% the sets $N_{i+1}$ as follows: if 
% $\forall x (\mA^{(i+1)}_1(x)\implies\next\mB^{(i+1)}_1(x))$ and
% $\forall x (\mA^{(i+1)}_2(x)\implies\next\mB^{(i+1)}_2(x))$ are full merged step clauses from $N_{i+1}$
% such that $\forall x (\mA^{(i+1)}_1(x)\implies\mA^{(i+1)}_2(x))$, we can delete 
% the first clause from $N_{i+1}$ yielding a formula $H_{i+1}'(x)$ equivalent
% to the original formula $H_{i+1}(x)$. We can also employ a weaker 
% \emph{subsumption test} instead of the stronger but computationally heavier
% implication test.
\end{note}
\begin{example}
Let us consider an unsatisfiable monodic temporal problem, $\TProb$, given by 
$$
\begin{array}{lcl}
\iP & = & \{\exists x A(x)\},\\
\uP & = & \{\forall x (B(x)\implies A(x)\land \lnot L(x))\},\\
\sP & = & \{ A(x)\implies\next B(x)\}, \\
\eP & = & \{\sometime L(x)\}
\end{array}
$$
and apply the BFS algorithm to $\sometime L(x)$.

The set of all full merged step clauses, $N_1$, whose right-hand sides imply
$\lnot L(x)$, is:
\begin{eqnarray}
(\forall y A(y))            & \implies & \next (\forall y B(y)),\label{eq:1}\\
(A(x) \land \forall y A(y)) & \implies & \next (B(x) \land \forall y B(y)),\label{eq:2}\\
(A(x) \land \exists y A(y)) & \implies & \next (B(x) \land \exists y B(y))\label{eq:3}.
\end{eqnarray}
Note that $\forall x (\forall y A(y) \implies A(x) \land \forall y A(y))$ and
$\forall x (A(x) \land \forall y A(y))\implies A(x) \land \exists y A(y))$; 
therefore, clauses $(\ref{eq:1})$ and $(\ref{eq:2})$ can be deleted from 
$N_1$ yielding 
$$
N_1' = \{(A(x) \land \exists y A(y))  \implies  \next (B(x) \land \exists y B(y))\}
\quad\text{ and }\quad
H_1'(x) = (A(x) \land \exists y A(y)).
$$

The set of all full merged step clauses $N_2$ whose right-hand sides imply
$L(x) \land H_1'(x)$ coincides with $N_1$ and the output of the 
algorithm is $H_2'(x) \equiv H_1'(x)$. 
The conclusion of the eventuality resolution rule, 
$\forall x \lnot A(x)\lor \lnot \exists y A(y)$, simplified to 
$\forall x \lnot A(x)$, contradicts the initial part of the problem.

Note that all full merged step clauses from $N_1$ are loops in $\sometime L(x)$,
but both conclusions of the eventuality resolution rule, applied to the loops
$(\ref{eq:1})$ and $(\ref{eq:2})$, can be simplified to 
$\exists x \lnot A(x)$ which does not contradict the initial part.
\end{example}

\section{Semantics with expanding domains}
\label{sec:expanding}
So far, we have been considering temporal formulae interpreted over
models with the \emph{constant domain assumption}. In this section we
consider another important case, namely models that have
\emph{expanding domains}. Although it is known that satisfiability
over expanding domains can be reduced to satisfiability over constant
domains~\cite{WZ01DecModal}, we here provide a procedure that can be
applied directly to expanding domain problems. Our interest in such
problems is partly motivated by the fact that the expanding domain
assumption leads to a simpler calculus, more amenable to practical
implementation~\cite{KDDFH03}, and partly by the correspondence
between expanding domain problems and important applications, such as
spatio-temporal logics~\cite{WZ02,GKKWZ03} and temporal description
logics~\cite{AFWZ02}. In addition, the way we refine the calculus of
Section~\ref{sec:ccalc} to the expanding domain case constitutes, we
believe, an elegant and significant simplification.
\smallskip

\noindent We begin by presenting the expanding domain semantics and
proceed to give the give the resolution calculus for the expanding
domain case.

\medskip

%We consider temporal models over \emph{expanding domains}, and
%
Under expanding domain semantics, formulae of \fotl are interpreted in
first-order temporal structures of the form $\gM = \langle D_n, I_n
\rangle$, $n\in\Nat$, where every $D_n$ is a non-empty set such that
whenever $n<m$, $D_n\subseteq D_m$,
%, the \emph{domain} of $\gM$, 
and $I_n$ is  an interpretation of predicate and constant symbols over $D_n$.
Again, we require that the interpretation of constants is rigid.
%Thus, for
%every constant $c$, and all moments of time $i,j\geq 0$, we have $I_i(c)=
%I_j(c)$.
A {(variable)
assignment} $\ga$ is a function from the set of individual
variables to $\cup_{n\in\Nat} D_n$;  the set of all assignments is denoted 
by $\gA$.

For every moment of time $n$, the corresponding {first-order}
structure, $\gM_n = \langle D_n, I_n\rangle$; the corresponding set of
variable assignments $\gA_n$ is a subset of the set of all assignments,
$\gA_n = \{\ga\in\gA\;|\; \ga(x)\in D_n \textrm{ for every variable $x$}\}$; 
clearly, $\gA_n\subseteq\gA_m$ if $n<m$.

Then, the \emph{truth} relation $\gM_n\models^\ga \phi$ in a structure $\gM$
is defined inductively in the same way as in the constant domain case, but
\emph{only for those assignments $\ga$ that satisfy the condition 
$\ga\in\gA_n$}.
%, is defined inductively  in the
%usual way under the following understanding of temporal operators:
%\texttt{Semantics goes here}

\medskip
\begin{example}\label{ex:cVSe} The formula 
$\forall x P(x)\land 
\always(\forall x P(x)\implies\next\forall x P(x))\land\sometime\exists y\lnot P(y)$ 
is unsatisfiable over both expanding and constant domains;
the formula
$\forall x P(x)\land 
\always(\forall x (P(x)\implies\next P(x)))\land\sometime\exists y \lnot P(y)$ 
is unsatisfiable over constant domains but has a model with an expanding
domain.
\end{example}
It can be seen that our earlier reduction to DSNF holds for the
expanding domain case (the only difficulty is
Lemma~\ref{th:unconditioning} where, in defining $\wfL(d)$, we must
consider cases where $\gM_k \models \always \sometime P(d)$ or $\gM_k
\models \sometime \always \notl P(d)$ where $k$ is the moment when $d$
``appears'').

The calculus itself coincides with the calculus given in
Section~\ref{sec:ccalc}; the only difference occurs in the merging
operation.  As Example~\ref{ex:cVSe} shows, the derived step clause
$(\ref{eq:forall})$ is not a logical consequence of
$(\ref{eq:premises})$ in the expanding domain case. Surprisingly, if
we omit derived step clauses of this form, we not only obtain a
correct calculus, but also a complete calculus for the expanding
domain case!

\begin{definition}[\Grounded{} Step Clauses: Expanding Domains]
Let $\TProb$ be a monodic temporal problem, and let
$$
P_{i_1}(x)\implies\next M_{i_1}(x), \dots, P_{i_k}(x)\implies\next M_{i_k}(x)
$$
be a subset of the set of its original non-ground step clauses. Then
$$
\begin{array}{c}
\exists x (P_{i_1}(x)\land\dots\land P_{i_k}(x))\implies\next\exists x (M_{i_1}(
x)\land\dots\land M_{i_k}(x)),\\
P_{i_j}(c)\implies\next M_{i_j}(c)
\end{array}
$$
are \emph{e-\grounded} step clauses, where $c$ is a constant occurring in
$\TProb$.
\end{definition}
The notions of a merged derived and full step clause as well as the 
calculus itself are exactly the same as in Section~\ref{sec:ccalc}.

Correctness of this calculus is again straightforward. 
As for completeness, we have to slightly modify the proof of
Section~\ref{sec:proof}.

The proof of Theorem~\ref{th:complete} 
relies on the theorem on existence of a model, Theorem~\ref{th:model}, and
it can be seen that if we prove an analog of Theorem~\ref{th:model} for the
expanding domain case, the given proof of completeness holds for the
this case.

We outline here how to modify the proof of Theorem~\ref{th:model} for
the case of expanding domains. All the definitions and properties
from Section~\ref{sec:proof} are transfered here with the following
exceptions.
\smallskip

\noindent Now, the universally quantified part does not contribute
either to $\mA$ or $\mB$. 
$$
\begin{array}{l}
\mathcal{A}_{\mathcal{C}} = \bigwedge\limits_{\gamma \in \Gamma} 
\exists x A_{\gamma}(x)  \andl A_{\theta} \andl
\bigwedge\limits_{c \in C} A_{\rho(c)}(c),\\
\mathcal{B}_{\mathcal{C}} = \bigwedge\limits_{\gamma \in \Gamma} 
\exists x B_{\gamma}(x)  \andl B_{\theta} \andl
\bigwedge\limits_{c \in C} B_{\rho(c)}(c). 
\end{array}
$$
This change affects the  suitability of predicate colours.
\begin{lemma}[Analogue of Lemma~\ref{lem:lemma3}]\label{lem:lemma3e}
Let $H$ be the behaviour graph for the problem
$\TProb =\, \langle\uP, \iP, \sP, \eP\rangle$ with an edge from a vertex
$\mC=(\Gamma, \theta,\rho)$ to a vertex $\mC'=(\Gamma', \theta',\rho')$.
Then 
\begin{enumerate}
\item[1.] for every $\gamma\in\Gamma$ there exists a $\gamma\,'\in\Gamma'$ such that
the pair $(\gamma,\gamma\,')$ is suitable;
\item[3.] the pair of propositional colours $(\theta,\theta')$ is suitable;
\item[4.] the pair of constant distributions $(\rho,\rho')$ is suitable.
\end{enumerate}
\end{lemma}
Note that the missing condition $2.$ of Lemma~\ref{lem:lemma3} does
not hold in the expanding domain case. However, under the conditions
of Lemma~\ref{lem:lemma3e}, if $\gamma'=\rho'(c)$,
for some $c\in\const(\TProb)$,
there always exists a $\gamma\in\Gamma$ such that the pair
$(\gamma,\gamma')$ is suitable.

Since for a predicate colour $\gamma$ there
may not exist a colour $\gamma'$ such that the pair $(\gamma',\gamma)$
is suitable, the notion of a run is reformulated.
\begin{definition}[Run]
Let $\pi$ be a path through a behaviour graph $H$ of a temporal problem
$\TProb$. By a \emph{run} in $\pi$ we mean a function $r(n)$ mapping its
domain, $\dom r = \{n\in\Nat\;|\; n\geq n_0\}$ for some $n_0\in\Nat$, to
$\bigcup_{i \in \Nat} \Gamma_i$ such that for every $n \in \dom r$, $r(n) \in
\Gamma_n$, $r(n)$ the pair $(r(n),r(n+1))$ is suitable. 
\end{definition}
\smallskip

\noindent Finally, the proof of Lemma~\ref{lem:path} is modified as
follows.
\begin{proof}[of Lemma~\ref{lem:path} for the expanding domain case]
We construct a path, $\pi$, through the behaviour graph, $H$,
satisfying properties (a), (b), and (d) in exactly the same way as in
the proof for constant domains. The only difference is in the way how
we prove condition (c).  We assume the denotation from that proof.
So, let $\mC = \pi(i)$ and $\gamma\in\Gamma_{\mC}$.  

Let $\mC = \pi(i)$ and $\gamma\in\Gamma_{\mC}$. Then there exists
$\gamma''\in\mC_n$
such that
$(\mC,\gamma)\to^+(\mC_{n},\gamma\,'')$. Since for every
$\gamma\,''\in\mC_n$ there exists $\gamma\,'''\in\mC_n^{(\gamma_{s_n},
L_k)}$ such that all eventualities are satisfied on the run-segment
from $\gamma\,''$ to $\gamma\,'''$ and there exists
$\gamma^{(4)}\in\mC_n$, $(\mC_n^{(\gamma_{s_n}, L_k)},
\gamma\,''')\to^+(\mC_n,\gamma^{(4)})$, then there is an e-run, $r$,
such that $r(i) = \gamma$, i.e., property (c) holds\footnote{We do not
assume any more that the e-run starts at $\mC_0$.}.
\end{proof}
This contributes to the following theorem.
\begin{theorem}[Correctness and Completness of Temporal Resolution for the
Expanding Domain Case]
The rules of temporal resolution preserve satisfiability.
Let an arbitrary monodic temporal problem $\TProb$ be
unsatisfiable over expanding domain.  Then there exists a successfully
terminating derivation by temporal resolution from $\TProb^c$.  
\end{theorem}

\section{Conclusions}
\label{sec:concl}
In this paper, we have modified and extended the clausal temporal
resolution technique in order to enable its use in monodic \fotl{}. We
have developed a specific normal form for \fotl{} and have provided a
complete resolution calculus for formulae in this form. The use of
this technique has provided us with increased understanding of the
monodic fragment, allowing definitions of new decidable monodic
classes, simplification of existing monodic classes by reductions, and
completeness of clausal temporal resolution in the case of monodic
logics with expanding domains.

However, not only is this approach useful in examining and extending
the monodic fragment, but it is being used as the basis for a
practical proof technique for certain monodic
classes~\cite{KDDFH03}. Refining and analyzing this implementation
forms part of our future work, as does the application of this
technique to a range of areas, including program verification,
temporal description logics, agent theories and spatio-temporal
logics.
\medskip

\paragraph{Acknowledgements}
The authors would like to acknowledge support from EPSRC via research
grants GR/M46631 and GR/R45376.

%
%\bibliography{main}

\end{document}